\documentclass[aps,prd,showpacs,eqsecnum]{revtex4}
\usepackage{amsmath,amssymb}
\usepackage{graphicx}

\begin{document}

\title{Relativistic stars with purely toroidal magnetic fields}

\author{Kenta Kiuchi$^{1}$
\footnote{\affiliation\ kiuchi@gravity.phys.waseda.ac.jp}}
\author{Shijun Yoshida$^{2}$
\footnote{\affiliation\ yoshida@astr.tohoku.ac.jp}}
\affiliation{$^{1}$Department of Physics, Waseda University, 3-4-1 Okubo,
 Shinjuku-ku, Tokyo 169-8555, Japan~}
\affiliation{$^{2}$Astronomical Institute, Tohoku University, Sendai 980-8578, Japan~}

\date{\today}

\begin{abstract}
We investigate the effects of the purely toroidal magnetic
field on the equilibrium structures of the relativistic stars.
The master equations for obtaining equilibrium solutions of relativistic
rotating stars containing purely toroidal magnetic fields are derived
for the first time. To solve these master equations numerically, we 
extend the Cook-Shapiro-Teukolsky scheme for calculating relativistic rotating
stars containing no magnetic field to incorporate the effects of the purely
toroidal magnetic fields. By using the numerical scheme, we then calculate
a large number of the equilibrium configurations for 
a particular distribution of  
the magnetic field in order to explore the equilibrium properties.
We also construct the equilibrium sequences of the constant baryon mass
and/or the constant magnetic flux, which model the evolution of an isolated neutron
star as it loses angular momentum via the gravitational waves.
Important properties of the equilibrium configurations of the magnetized stars
obtained in this study are summarized as follows ; (1) For the non-rotating stars,
the matter distribution of the stars is prolately distorted due to
the toroidal magnetic fields. (2) For the rapidly rotating
stars, the shape of the stellar surface becomes oblate because
of the centrifugal force. But, the matter distribution deep inside the star is
sufficiently prolate for the mean matter distribution of the star to be prolate.
(3) The stronger toroidal magnetic fields lead to the mass-shedding of the stars
at the lower angular velocity.
(4) For some equilibrium sequences of the constant baryon mass and
magnetic flux, the stars can spin up as they lose angular momentum.
\end{abstract}


\maketitle

\baselineskip 18pt

%**************************************
\section{Introduction}\label{sec:Intro}
%**************************************
There is growing evidence for the existence of so-called magnetars, 
supermagnetized neutron stars with the magnetic fields of 
$\sim 10^{14}-10^{15}$ G (e.g., \cite{Lattimer:2006,wod}). 
Although such stars are estimated to be only a subclass 
($\sim 10 \%$) of the canonical neutron stars 
with $\sim 10^{12}-10^{13}$ G~\cite{kouve}, much attention has
been paid to the objects because they pose many
astrophysically exciting but unresolved problems.
Giant flaring activities in the soft gamma repeaters observed 
for the last two decades have given us good opportunities to study 
the coupling of the interior to the magnetospheric structures of magnetars 
\cite{Thompson:1995gw,Thompson:1996pe}, 
but the relationship between the crustal fraction and the subsequent
star quarks has not yet been clarified (see references in \cite{anna,
Geppert:2006cp}).
The origin of the large magnetic field is also a big problem, whether
generated at post-collapse in the rapidly rotating neutron star 
\cite{Thompson:1993hn} 
or descended from the main sequence stars \cite{Ferrario:2007bt}.
Assuming large magnetic fields before core-collapse, 
researches have recently performed extensive magnetohydrodynamic (MHD) stellar collapse
simulations~\cite{Yamada:2004,Kotake:2004,Moissenko:2006,Obergaulinger:2006,
Shibata:2006hr,Livne:2007}
in order to understand the formation mechanism of
magnetars, with different levels of sophistication in the
 treatment of equations of state (EOSs), the neutrino transport,
 and general relativity (see \cite{Kotake:2006} 
for a review). 
Here it is worth
mentioning that the gravitational waves from magnetars could be a
useful source of information about magnetar interiors 
~\cite{Bonazzola:1995rb,Cutler:2002nw}.
 From a microscopic point of view, 
the effects of magnetic field larger than the so-called QED 
limit of $\sim 4.4\times
 10^{13}~\rm{G}$, on the EOSs (e.g., \cite{Lattimer:2006}
) and 
the radiation processes 
have been also extensively investigated (see \cite{Harding:2006} 
for a review). 
For the understanding
of the formation and evolution of the magnetars, it is indispensable to
unify these macroscopic and microscopic studies, albeit 
not an easy job.

When one tries to study the above interesting issues,  
the construction of an equilibrium configuration of magnetars 
may be one of the most fundamental problems. 
In spite of the many magnetized star investigations that have so far 
been carried out, with different levels of sophistication in the 
treatment of the magnetic field structure, equations of state (EOSs), 
and the general relativity~\cite{Bocquet:1995,Ioka:2003,
Ioka:2004,Kiuchi:2007pa,Konno:1999zv,
Tomimura:2005,Yoshida:2006a,Yoshida:2006b}, there still remains room 
for more sophistication in the previous studies in which stellar structures 
are obtained within the framework of Newtonian dynamics. Since 
we are interested in magnetar structures, the general relativistic 
treatment due to the strong gravity of magnetars is required 
in constructions of a magnetar model. However, a fully general relativistic 
means constructing a stellar model with arbitrarily magnetic structure 
are still not available. In most studies on the structures of 
relativistic magnetized stars, weak magnetic fields and/or purely poloidal 
magnetic fields have been assumed. 
Bocquet et al.~\cite{Bocquet:1995} and Cardall et al.~\cite{Cardall:2001} have 
treated relativistic stellar models containing purely poloidal magnetic fields. 
Konno et al.~\cite{Konno:1999zv} have analyzed similar models using 
a perturbative approach. 
Ioka and Sasaki have investigated structures of mixed poloidal-toroidal 
magnetic fields around a spherical star by using a perturbative 
technique~\cite{Ioka:2004}.

As shown in core-collapse supernova MHD simulations (see, e.g., \cite{Kotake:2006}), 
the toroidal magnetic fields can be easily amplified because of 
the winding up of the initial seed poloidal fields as long as the core is rotating 
differentially. After core bounce, as a result, 
the toroidal fields dominate over the poloidal ones in 
some proto-neutron star models even if there is no toroidal field initially. This 
winding up amplification of the toroidal magnetic fields therefore indicates 
that some neutron stars could have toroidal fields much higher than poloidal ones. 
As mentioned before, however, no effect of a strong toroidal magnetic field has been 
taken into account in relativistic magnetized stellar models. 
In order to elucidate the effects of strong toroidal magnetic fields on neutron star 
structures, in this paper, we concentrate on magnetic effects due to 
purely toroidal fields on structures of relativistic stars as a first step. 
Within the framework of the Newtonian dynamics, Miketinac has obtained 
highly magnetized stars containing purely toroidal magnetic fields~\cite{Miketinac:1973}. 
In this study, we extend Miketinac's models so as to include the effects of general 
relativity and rotation.

First of all, we derive master equations 
for obtaining relativistic rotating stellar models having purely toroidal magnetic 
fields. The master equations obtained are converted into integral expressions and solved 
numerically with a self-consistent field scheme similar to the Komatsu-Eriguchi-Hachisu 
scheme for equilibrium structures of relativistic rotating stars~\cite{Komatsu:1989}. 
Using the numerical 
scheme written in the present study, we construct a large number of equilibrium stars 
for a wide range of parameters.  As shown by Trehan \& Uberoi~\cite{trehan1972}, the toroidal 
magnetic field tends to distort a stellar shape prolately. This is because the toroidal 
magnetic field lines behave like a rubber belt, pulling in the matter around 
the magnetic axis. Such a prolate-shaped neutron star may be an optimal source of 
gravitational waves if the star rotates~\cite{Cutler:2002nw}. In this gravitational 
wave emission mechanism, one important parameter is degree of prolateness, 
which determines the amount of gravitational radiation. Thus, we pay attention 
to the deformation of the star. 
We also construct sequences of the equilibrium stars along which the total baryon rest 
mass and/or the magnetic flux with respect to the meridional cross-section keep constant.  
These equilibrium sequences are available for exploring evolutionary sequences because 
the total baryon rest mass and the magnetic flux are well conserved in quasi-steady 
evolution of isolated neutron stars.

This paper is organized as follows. The master equations for obtaining equilibrium 
configurations of rotating stars with purely toroidal magnetic fields are derived 
in Sec.~\ref{sec:basic}. The numerical scheme for computing the stellar models is 
briefly described in Sec.~\ref{sec:numsch}. Sec.~\ref{sec:result} is devoted to 
showing numerical results. Summary and discussion follow in Sec.~\ref{sec:summary}.
In this paper, we use geometrical units with $G=c=1$.

%*****************************************
\section{Master equations}\label{sec:basic}
%*****************************************

%*****************************************************************
\subsection{Matter equations}\label{subsec:Bernol}
%*****************************************************************

The rotating relativistic stars containing purely toroidal magnetic fields are 
considered in this paper. Assumptions to obtain the equilibrium 
models are summarized as follows ;  
(1) Equilibrium models are stationary and axisymmetric, i.e. the spacetime has the time Killing 
vector, $t^\mu$, and the rotational Killing vector, $\varphi^\mu$, and Lie derivatives of 
the physical quantities of the equilibrium models with respect to these Killing vectors vanish. 
(2) The matter source is approximated by a perfect fluid with infinite conductivity. 
(3) There is no meridional flow of the matter. (4) The equation of state for the matter is barotropic. 
By virtue of assumption (3), the matter four-velocity is reduced to
%---------------------------------------------------------------
\begin{eqnarray}
u^\mu  = u^0 ( t^\mu + \Omega \varphi^\mu ), \label{eq:4-vel}
\end{eqnarray}
%---------------------------------------------------------------
where $\Omega$ is the angular velocity of the matter. We further assume 
that the magnetic field of the star has toroidal parts only, 
so that the Faraday tensor, $F_{\mu\nu}$, obeys the conditions, given by   
%---------------------------------------------------------------
\begin{eqnarray}
F_{\mu\nu}\,t^\nu=0\,, \quad  F_{\mu\nu}\,\varphi^\nu=0\,. \label{eq:Fraday}
\end{eqnarray}
%---------------------------------------------------------------
Note that since Eq.~(\ref{eq:Fraday}) includes five independent conditions,  
we see that the Faraday tensor has one independent component only. 
Thanks to the conditions assumed above, one of  the Maxwell equations and the perfect-conductivity condition,  
%--------------------------------------------
\begin{eqnarray}
&& F_{[\mu\nu,\alpha]} = 0,\label{eq:Maxwel}\\
&& F_{\mu\nu}u^\nu     = 0,\label{eq:MHD}
\end{eqnarray}
%---------------------------------------------
are automatically satisfied, and we need not consider them any further. 
The stress-energy tensor of the perfectly conductive fluid is given by 
%------------------------------------------------------------------------
\begin{eqnarray}
T^{\mu\nu} = ( \rho_0 + \rho_0 e + P ) u^\mu u^\nu + P g^{\mu\nu} 
+ \frac{1}{4\pi}\left[B^\alpha B_\alpha 
\left( u^\mu u^\nu + \frac{1}{2}g^{\mu\nu} \right) - B^\mu B^\nu \right],
\label{eq:stress-tensor}
\end{eqnarray}
%-------------------------------------------------------------------------
where $\rho_0$ is the rest energy density, $e$ the specific internal 
energy, $P$ the pressure, and $g^{\mu\nu}$ the inverse of the metric $g_{\mu\nu}$. 
Here, we have defined the magnetic field 
measured by the observer with the fluid four-velocity $u^\mu$ as 
%-----------------------------------------------------------------------
\begin{eqnarray}
B^\mu = - \frac{1}{2} \epsilon^{\mu\nu\alpha\beta}u_\nu F_{\alpha\beta}, 
\label{eq:Bfield}
\end{eqnarray}
%-----------------------------------------------------------------------
where $\epsilon^{\mu\nu\alpha\beta}$ stands for the contravariant Levi-Civita tensor 
with $\epsilon^{0123}=-1/\sqrt{-g}$. Here $g$ denotes the determinant of the metric. 
By using  Eqs.~(\ref{eq:4-vel})--(\ref{eq:Fraday}), (\ref{eq:MHD})--(\ref{eq:stress-tensor}), 
and (\ref{eq:Bfield}), we can prove that the circularity condition, 
 i.e., $t_\alpha T^{\alpha[\beta} t^\gamma \varphi^{\delta]} = 
\varphi_\alpha T^{\alpha[\beta} t^\gamma \varphi^{\delta]} = 0$, is satisfied in the 
present situation~\cite{Wald:1984}. Note that a similar argument about the circularity has been given by 
Oron~\cite{Oron: 2002}. 
Therefore, there exists a family of two-surfaces everywhere 
orthogonal to the plane 
defined by the two Killing vectors $t^\mu$ and $\varphi^\mu$~\cite{Carter:1969}, 
and the metric can be written, following~\cite{Komatsu:1989,Cook:1992}
, in the form

\begin{eqnarray}
ds^2 = - {\rm e}^{\gamma+\rho} dt^2 + {\rm e}^{2\alpha}( dr^2 + r^2 d\theta^2 ) 
+ {\rm e}^{\gamma-\rho}r^2\sin^2\theta( d\varphi -\omega dt )^2,\label{eq:metric}
\end{eqnarray}

where the metric potentials, $\gamma$, $\rho$, $\alpha$, and $\omega$, 
are functions of $r$ and $\theta$ only. We see that the non-zero component of $F_{\mu\nu}$ 
in this coordinate is $F_{12}$. It is convenient to define the determinants of 
two two-dimensional subspaces for later discussion, 

\begin{eqnarray}
g_A &\equiv& g_{11} g_{22}-( g_{12} )^2={\rm e}^{4\alpha}\,r^2 ,~~\nonumber \\
g_B &\equiv& - g_{00} g_{33} + ( g_{03} )^2={\rm e}^{2\gamma}\, r^2\sin^2\theta  
,\label{eq:det}
\end{eqnarray}

where the determinant of the metric~(\ref{eq:metric}) is related 
to the two determinants as $g=-g_A g_B$. We then find a convenient relation, 

\begin{eqnarray}
F_{12} = g_A F^{12}.\label{eq:Fcon}
\end{eqnarray}

Because of this relation, the four-current $J^\mu$ can, in terms of $F_{12}$,  
be written in the form,
\begin{eqnarray}
J^\alpha 
&=& 
\frac{1}{4\pi} \frac{1}{\sqrt{-g}}\partial_\beta
(\sqrt{-g}F^{\alpha\beta}) \nonumber\\
&=&
\frac{1}{4\pi} \frac{1}{\sqrt{g_Ag_B}}
\left[
- \partial_1 \left( \sqrt{\frac{g_B}{g_A}} F_{12} \right) \delta^{\alpha}_2
+ \partial_2 \left( \sqrt{\frac{g_B}{g_A}} F_{12} \right) \delta^{\alpha}_1
\right].\label{eq:4-curr}
\end{eqnarray}

The relativistic Euler equation, 

\begin{eqnarray}
u^\nu \nabla_\nu u^\mu + \frac{1}{\rho_0 h}\nabla_\mu P 
- \frac{1}{\rho_0 h}F_{\mu\nu} J^\nu = 0, \label{eq:Euler}
\end{eqnarray}

is reduced to 

\begin{eqnarray}
- \partial_A \ln u^0  + \frac{1}{\rho_0 h} \partial_A P
+ \frac{1}{4\pi\rho_0 hg_B}\sqrt{\frac{g_B}{g_A}}F_{12}
\partial_A \left( \sqrt{\frac{g_B}{g_A}}F_{12} \right ) = 0,\label{eq:Euler2}
\end{eqnarray}

where $h=1+e+P/\rho_0$ is the specific enthalpy and we have assumed the uniform  
rotation, i.e., $\Omega={\rm constant}$. Note that although the extension to 
the case of the differential rotation is straightforward, we take the uniform 
rotation law for the sake of simplicity. Integrability of Eq.~(\ref{eq:Euler2}) requires

\begin{eqnarray}
\sqrt{\frac{g_B}{g_A}}F_{12} = K(u);~~~u \equiv \rho_0 hg_B,
\label{eq:I-con1}
\end{eqnarray}

where $K$ is an arbitrary function of $\rho_0 hg_B$. 
Integrating Eq.~(\ref{eq:Euler2}), we arrive at the equation of hydrostatic equilibrium, 

\begin{eqnarray}
\ln h - \ln u^0
+ \frac{1}{4\pi} \int \frac{K(u)}{u}\frac{dK}{du}du = C,\label{eq:Ber}
\end{eqnarray}

where $C$ is an integral constant. 
The nontrivial components of the magnetic field are explicitly given by 

\begin{eqnarray}
&& B_0 = - \Omega u^0 K(u),\label{eq:comp-mag1}\\
&& B_3 =          u^0 K(u), \label{eq:comp-mag2}
\end{eqnarray}

in which we confirm that our definition of the toroidal magnetic field (\ref{eq:Fraday}) reasonably 
agrees with the Newtonian definition of the toroidal field.

With the metric~(\ref{eq:metric}), the four-velocity~(\ref{eq:4-vel}), 
the proper velocity of the matter with respect to 
a zero angular momentum observer (ZAMO), $v$, 
the components of the magnetic field~(\ref{eq:comp-mag1})-(\ref{eq:comp-mag2}), 
and the hydrostatic equilibrium equation~(\ref{eq:Ber})
can be, respectively, written as 
%-----------------------------------------------------------------------------
\begin{eqnarray}
&&
u^\mu = \frac{{\rm e}^{-(\rho+\gamma)/2}}{\sqrt{1-v^2}}
( t^\mu + \Omega \varphi^\mu ),\label{eq:4-vel2}\\
&&
v = ( \Omega - \omega )r\sin\theta {\rm e}^{-\rho}.\label{eq:p-vel}\\
&& B_0 = -\frac{{\rm e}^{-(\rho+\gamma)/2}}{{\sqrt{1-v^2}}}\Omega K(u),\label{eq:B0}\\
&& B_3 =  \frac{{\rm e}^{-(\rho+\gamma)/2}}{{\sqrt{1-v^2}}}       K(u),\label{eq:B3}
\\
&& \ln h + \frac{\gamma+\rho}{2} + \frac{1}{2} \ln ( 1 - v^2 )
+ \frac{1}{4\pi} \int \frac{K(u)}{u}\frac{dK}{du}du = C.\label{eq:Ber2}
\end{eqnarray}
%-----------------------------------------------------------------------------

%***********************************************
\subsection{Einstein equations}\label{subsec:Ein}
%***********************************************
Following~\cite{Komatsu:1989,Cook:1992}, we may write the general relativistic field 
equations determining $\rho$, $\gamma$ , and $\omega$ as 
%--------------------------------------------------------------------------
\begin{eqnarray}
&& \nabla^2 [\rho {\rm e}^{\gamma/2} ] = S_\rho(r,\mu),\label{eq:basic-rho}\\
&& \left( \nabla^2 + \frac{1}{r}\partial_r - \frac{\mu}{r^2}\partial_\mu 
\right) [ \gamma {\rm e}^{\gamma/2} ] = S_\gamma (r,\mu),\label{eq:basic-gam}\\
&& \left( \nabla^2 + \frac{2}{r}\partial_r - \frac{2\mu}{r^2}\partial_\mu 
\right) [\omega {\rm e}^{(\gamma-2\rho)/2} ] = S_\omega (r,\mu),
\label{eq:basic-omg}
\end{eqnarray}
%---------------------------------------------------------------------------
where $\nabla^2$ is the Laplacian of flat 3-space, and 
$\mu=\cos\theta$. Here, $S_\rho$, $S_\gamma$, and $S_\omega$ are given by 
%----------------------------------------------------------------------------
\begin{eqnarray}
&& S_\rho(r,\mu) = {\rm e}^{\gamma/2} \Big \{
8 \pi {\rm e}^{2\alpha} ( \rho_0 + \rho_0 e + P )\frac{1+v^2}{1-v^2} + 
r^2 ( 1 - \mu^2 ) {\rm e}^{-2\rho} \left( \omega_{,r}^2 + \frac{1-\mu^2}{r^2}
\omega_{,\mu}^2 \right) \nonumber\\
&& + \frac{1}{r}\gamma_{,r} - \frac{\mu}{r^2}\gamma_{,\mu} 
+ \frac{\rho}{2} \left[16 \pi {\rm e}^{2\alpha} \left( P + \frac{|B|^2}{8\pi} 
\right)-\gamma_{,r} \left( \frac{1}{2} \gamma_{,r} + \frac{1}{r} 
\right ) - \frac{1}{r^2}\gamma_{,\mu} \left( 
\frac{1-\mu^2}{2}\gamma_{,\mu} - \mu \right)\right]
\Big \},\label{eq:soc-rho}\\
&& S_\gamma(r,\mu) = {\rm e}^{\gamma/2} \left\{
16 \pi {\rm e}^{2\alpha} \left( P + \frac{|B|^2}{8\pi} \right)
+ \frac{\gamma}{2} 
\left[ 16 \pi {\rm e}^{2\alpha} \left( P + \frac{|B|^2}{8\pi} \right) 
- \frac{1}{2}\gamma_{,r}^2 - \frac{1-\mu^2}{2r^2}\gamma_{,\mu}^2\right]
\right \},\label{eq:soc-gam}\\
&& S_\omega(r,\mu) = {\rm e}^{(\gamma-2\rho)/2} \Big\{
- 16\pi {\rm e}^{2\alpha} \frac{(\Omega-\omega)(\rho_0 + \rho_0 e + P )}{1-v^2}
\nonumber\\
&& + \omega \Big[ 
- 8 \pi {\rm e}^{2\alpha} \left( 
\frac{(\rho_0 + \rho_0 e + P ) ( 1 + v^2 )}{1-v^2}
- P - \frac{|B|^2}{8\pi}
\right) 
- \frac{1}{r} 
\left( 2 \rho_{,r} + \frac{1}{2}\gamma_{,r}\right)
+ \frac{\mu}{r^2} \left( 2 \rho_{,\mu} + \frac{1}{2} \gamma_{,\mu} \right)
\nonumber\\
&& + \frac{1}{4}( 4\rho_{,r}^2 - \gamma_{,r}^2 )
   + \frac{1-\mu^2}{4r^2} ( 4\rho_{,\mu}^2 - \gamma_{,\mu}^2 )
   - r^2 ( 1 - \mu^2 ) {\rm e}^{-2\rho} \left( \omega_{,r}^2 
   + \frac{1-\mu^2}{r^2}\omega_{,\mu}^2 \right)
\Big]
\Big\}
,\label{eq:soc-omg}
\end{eqnarray}
where $|B|$ denotes the magnitude of the magnetic field, given by  

\begin{eqnarray}
|B| = \frac{|K|{\rm e}^{-\gamma}}{r\sin\theta}.\label{eq:mag-mag}
\end{eqnarray}

Since $S_\rho$, $S_\gamma$, and $S_\omega$ do not include the second-order 
derivative of the metric potentials, the three equations 
(\ref{eq:basic-rho})--(\ref{eq:basic-omg}) 
are all elliptic-type partial differential equations. We can solve these 
equations iteratively with appropriate boundary conditions by using 
a self-consistent field method~\cite{Komatsu:1989}. 

The fourth field equation determining $\alpha$ is given by

\begin{eqnarray}
&&\alpha_{,\mu} = - \frac{1}{2} ( \rho_{,\mu} + \gamma_{,\mu} ) 
- \{ ( 1 - \mu^2 )( 1 + r \gamma_{,r} )^2 + [ \mu 
- ( 1 -\mu^2 )\gamma_{,\mu}]^2 \}^{-1} 
\Big[
\nonumber\\
&&
\frac{1}{2} [ r^2 ( \gamma_{,rr} + \gamma_{,r}^2 ) - 
( 1 - \mu^2 ) ( \gamma_{,\mu\mu} + \gamma_{,\mu}^2 )]
[ - \mu + ( 1 - \mu^2 ) \gamma_{,\mu} ]
 + r \gamma_{,r} \left[
 \frac{1}{2}\mu + \mu r \gamma_{,r} 
 + \frac{1}{2}( 1 - \mu^2 ) \gamma_{,\mu} 
\right] 
\nonumber\\
&& 
+ \frac{3}{2}\gamma_{,\mu} [ - \mu^2 + \mu ( 1 - \mu^2 )\gamma_{,\mu}]
 - r ( 1 - \mu^2 ) ( \gamma_{,r\mu} + \gamma_{,r} \gamma_{,\mu} )
( 1 + r \gamma_{,r} ) 
- \frac{1}{4}\mu r^2 ( \rho_{,r} + \gamma_{,r} )^2  
\nonumber\\
&& 
- \frac{r}{2} ( 1 - \mu^2 ) ( \rho_{,r} + \gamma_{,r} ) 
( \rho_{,\mu} + \gamma_{,\mu} ) 
+ \frac{1}{4}\mu ( 1 - \mu^2 ) ( \rho_{,\mu}
+ \gamma_{,\mu} )^2 
- \frac{r^2}{2}( 1 -\mu^2)\gamma_{,r}(\rho_{,r} + \gamma_{,r} )
( \rho_{,\mu} + \gamma_{,\mu} ) 
\nonumber\\
&&
+ \frac{1}{4} ( 1 - \mu^2 ) \gamma_{,\mu}[
r^2 ( \rho_{,r} + \gamma_{,r} )^2 - ( 1 - \mu^2 )( \rho_{,\mu} 
+ \gamma_{,\mu} )^2]
+ ( 1 - \mu^2 ) e^{-2\rho} \Big\{
\frac{1}{4}r^4\mu \omega_{,r}^2 + \frac{1}{2} r^3 ( 1 - \mu^2 ) \omega_{,r}
\omega_{,\mu} 
\nonumber\\
&&
- \frac{1}{4}r^2 \mu ( 1 - \mu^2 )\omega_{,\mu}^2
+ \frac{1}{2}r^4 ( 1 - \mu^2 ) \gamma_{,r}\omega_{,r}\omega_{,\mu}
- \frac{1}{4}r^2 ( 1 - \mu^2 ) \gamma_{,\mu}
[ r^2 \omega_{,r}^2 - ( 1 - \mu^2 )\omega_{,\mu}^2 ]
\Big\}
\Big]\,.\label{eq:alp}
\end{eqnarray}

We can solve Eq.~(\ref{eq:alp}) by integrating it from the pole to the equator 
with an initial condition, given by 

\begin{equation}
\alpha={1\over 2}(\gamma-\rho) \quad {\rm at}\quad \mu=1\,. 
\end{equation}

%**************************************************
\subsection{Equation of state and  function $K(u)$}
\label{subsec:func}
%**************************************************
In this study, we adapt a polytropic equation of state 

\begin{eqnarray}
P = K_p \rho_0^{1+1/n},\label{eq:EOS}
\end{eqnarray}

where $n$ and $K_p$ are the polytropic index and constant, respectively. 
Because all the variables can be normalized by the factor $K_p^{n/2}$, whose unit 
is of length, we define the following dimensionless quantities,
\begin{eqnarray*}
\tilde{r} = K_p^{-n/2} r,~~
\tilde{\omega} = K_p^{n/2} \omega,~~
\tilde{\Omega} = K_p^{n/2} \Omega,~~
\tilde{\rho_0} = K_p^{n}   \rho_0,~~
\tilde{P}      = K_p^{n}   P,~~
\tilde{B}^\mu  = K_p^{n/2} B^\mu.
\end{eqnarray*}

For the arbitrary function $K$ defined in~(\ref{eq:I-con1}), we take the following 
simple form, 

\begin{eqnarray}
K(u) = b u^k,\label{eq:funcK}
\end{eqnarray}

where $b$ and $k$ are constants. Then, $|B|$ and $J^\alpha$ are reduced into 

\begin{eqnarray}
|B|=|b|\,(\rho_0 h)^k {\rm e}^{(2k-1)\gamma}({r\sin\theta})^{2k-1}
\,,  \label{eq:b}
\end{eqnarray}

\begin{eqnarray}
J^\alpha &=& \frac{b\,k\,{\rm e}^{(2k-3)\gamma}}{4\pi\, r\, {\rm e}^{2\alpha}}\,(\rho_0 h)^{k-1}  
(r\sin\theta)^{2k-3}
\left[\partial_\theta \left( \rho_0 h r^2\sin^2\theta {\rm e}^{2\gamma} \right) \delta^{\alpha}_1
- \partial_r \left( \rho_0 h r^2\sin^2\theta {\rm e}^{2\gamma} \right) \delta^{\alpha}_2
\right].\label{eq:4-curr2}
\end{eqnarray}

Regularity of $B^\alpha$ on the magnetic axis requires that $k \ge 1$.  
As shown in equation (\ref{eq:b}), the magnetic fields vanish at the surface of the star 
when $k \ge 1$, which we require from the regularity of $B^\alpha$ as mentioned above. 
Therefore, we need not pay attention to the boundary condition at the stellar surface 
for the magnetic field when $k\ge 1$.  

%*******************************************
\section{Numerical Method}\label{sec:numsch}
%*******************************************

%*************************************
\subsection{Cook-Shapiro-Teukolsky scheme}\label{sec:CST}
%*************************************

As shown in the last section, our master equations are quite similar to those 
for rotating relativistic stars in the sense that the type and rank of 
the differential equations are the same. The difference is the emergence of 
the new function $K(u)$, which can be treated in the same way as that of 
$\Omega$ for differentially rotating stars. In order to solve our master 
equations, therefore, we can straightforwardly adapt a numerical method for computing 
rotating relativistic stellar models. In this study, we extend the 
Cook-Shapiro-Teukolsky scheme~\cite{Cook:1992} to incorporate the effects of 
purely toroidal magnetic fields. The Cook-Shapiro-Teukolsky scheme is 
a variant of the Komatsu-Eriguchi-Hachisu scheme~\cite{Komatsu:1989}, which 
is a general relativistic version of the Hachisu Self-Consistent Field scheme
for computing Newtonian equilibrium models of rotating stars~\cite{Hachisu:1986}. 
In the Cook-Shapiro-Teukolsky scheme, a new radial coordinate $s\in [0,1]$, defined as 
%-----------------------------------------------------------------
\begin{eqnarray}
\tilde{r} = \tilde{r}_e \left( \frac{s}{1-s} \right),\label{eq:s}
\end{eqnarray}
%-----------------------------------------------------------------
where $\tilde{r}_e$ is the dimensionless radius of the surface at the equator and 
this transformation maps radial infinity to the finite coordinate locations $s=1$, 
is introduced so as to include all the information of spacetime on the hypersurface 
of $t=$constant. Since our numerical 
scheme is quite similar to the Cook-Shapiro-Teukolsky scheme, the details are 
omitted here.

%********************************
\subsection{Global physical quantities}
%********************************
After obtaining solutions, it is useful to compute global physical quantities characterizing 
the equilibrium configurations to clearly understand the properties of 
the sequences of the equilibrium models. In this paper, we compute the following 
quantities: the gravitational mass $M$, the baryon rest mass $M_0$, the proper mass 
$M_p$, the total angular momentum $J$, the total rotational energy $T$, the total 
magnetic energy $H$, 
the quadrupole momentum $I_{xx}$, $I_{zz}$, and the magnetic flux $\Phi$, defined as 

\begin{eqnarray}
&&
M 
= \int ( - 2 {T^0}_0 + {T^\alpha}_\alpha )\sqrt{-g}d^3x \nonumber\\
&&
= 2\pi\int
\left\{
2P + \frac{\epsilon+P}{1-v^2} 
(1+v^2+2{\rm e}^{-\rho}r\sin\theta v\omega )
+ \frac{|B|^2}{4\pi}
\right\}
{\rm e}^{2\alpha+\gamma}r^2\sin\theta
drd\theta, 
\label{eq:ADMmass}\\
&&
M_0 = \int \rho_0 u^0 \sqrt{-g} d^3x
= 2\pi \int \frac{\rho_0}{\sqrt{1-v^2}}{\rm e}^{2\alpha+(\gamma-\rho)/2}r^2\sin\theta
drd\theta 
\label{eq:bmass}\\
&&
M_p = \int \epsilon u^0 \sqrt{-g} d^3x 
= 2\pi
\int \frac{\epsilon}{\sqrt{1-v^2}}{\rm e}^{2\alpha+(\gamma-\rho)/2}r^2\sin\theta
drd\theta 
\label{eq:pmass}\\
&&
J = \int {T^0}_3 \sqrt{-g} d^3x
= 2\pi \int ( \epsilon + P )\frac{v}{1-v^2}{\rm e}^{2\alpha+\gamma-\rho}
r^3\sin^2\theta drd\theta
\label{eq:amom}\\
&&
T = \frac{1}{2} \int \Omega {T^0}_3 \sqrt{-g} d^3x
= \pi \int ( \epsilon + P )\frac{v\Omega}{1-v^2}
{\rm e}^{2\alpha+\gamma-\rho}r^3\sin^2\theta
drd\theta
\label{eq:rot-ene}\\
&&
H = \frac{1}{8\pi}
\int B^\alpha B_\alpha u^0 \sqrt{-g} d^3x
= \frac{1}{4}\int \frac{|B|^2}{\sqrt{1-v^2}}
{\rm e}^{2\alpha+(\gamma-\rho)/2}r^2\sin\theta
drd\theta
\label{eq:mag-ene}\\
&&
I_{xx} = \pi \int \epsilon r^4 \sin\theta(1+\cos^2\theta)drd\theta
\label{eq:Ixx}\\
&&
I_{zz} = 2\pi \int \epsilon r^4 \sin^3\theta drd\theta
\label{eq:Izz}\\
&&
\Phi = \int F_{12} drd\theta 
= \int \frac{{\rm e}^{2\alpha-\gamma}}{\sin\theta}
K drd\theta
\label{eq:mag-flux}\,.
\end{eqnarray}
%**********************************************************************
The definition of the magnetic energy and flux are given in 
appendix~\ref{sec:apdixA}.
The gravitational energy $W$ and the mean deformation rate $\bar{e}$
are, respectively, defined as  
\begin{eqnarray}
&&
|W| = M_p + T + H - M,\label{eq:bind-ene}\\
&&
\bar{e} = \frac{{I}_{zz}-{I}_{xx}}{{I}_{zz}}.\label{eq:deform-rate}
\end{eqnarray}
%******************************************
\subsection{Virial identities}
%******************************************
Bonazzola and Gourgoulhon have derived two different kinds of virial identities 
for relativistic equilibrium configurations~\cite{Gourgoulhon:1994,Bonazzola:1994}. 
These virial identities provide us means to estimate error involved in the numerical solutions 
of the equilibrium configurations. Let us define two quantities $\lambda_2$ and 
$\lambda_3$, given by 
%------------------------------------------------------------------
\begin{eqnarray}
&&
\lambda_2 = 8 \pi \int^\infty_0 \int^\pi_0
\left[
P + ( \epsilon + P )\frac{v^2}{1-v^2} 
- \frac{|B|^2}{8\pi}
\right]{\rm e}^{2\alpha}rdrd\theta
\nonumber\\
&&
\times
\Big\{
\int^\infty_0 \int^\pi_0
\left[
\frac{1}{4}\partial(\gamma+\rho)\partial(\gamma+\rho)
-\frac{3}{4}{\rm e}^{-2\rho}r^2\sin^2\theta \partial \omega \partial \omega
\right]rdrd\theta
\Big\}^{-1}\label{eq:rela-vir2}\\
&&
\lambda_3 = 4\pi\int^\infty_0\int^\pi_0
\left[
3 P + (\epsilon+P)\frac{v^2}{1-v^2}
+ 
\frac{|B|^2}{8\pi}
\right]{\rm e}^{2\alpha+(\gamma-\rho)/2}r^2\sin\theta dr d\theta
\nonumber\\
&&
\times
\Big\{
\int^\infty_0 \int^\pi_0 
\Big[
\partial(\gamma+\rho)\partial(\gamma+\rho)
-\partial\alpha\partial(\gamma-\rho)
\nonumber\\
&&
+ ( 1 - {\rm e}^{2\alpha-\gamma+\rho} ) \frac{1}{r}
\Big(
- 2 \alpha_{,r}
-   \frac{2\cot\theta}{r}\alpha_{,\theta}
+ \frac{1}{2}(\gamma-\rho)_{,r}
+ \frac{\cot\theta}{2r}(\gamma-\rho)_{,\theta}
\Big)
\nonumber\\
&&
-\frac{3}{2}{\rm e}^{-2\rho}r^2\sin^2\theta\partial\omega\partial\omega
\Big]{\rm e}^{(\gamma-\rho)/2}r^2\sin\theta dr d\theta
\Big\}^{-1},\label{eq:rela-vir3}
\end{eqnarray}

where $\partial \alpha \partial \beta$ is the abbreviation of

\begin{eqnarray}
\partial \alpha \partial \beta = \alpha_{,r} \beta_{,r} + \frac{1}{r^2}
\alpha_{,\theta} \beta_{,\theta}.
\label{eq:ab1}
\end{eqnarray}

Then we may define two error indicators, given by  

\begin{eqnarray}
&&\text{GRV2} \equiv |1-\lambda_2|\,,\label{eq:grav2}\\
&&\text{GRV3} \equiv |1-\lambda_3|\,,\label{eq:grav3}
\end{eqnarray}

which vanish if the virial identities are strictly satisfied. In all 
the numerical computations shown in the next section, GRV2 and GRV3 are monitored 
to check accuracy of the numerical solutions. A typical value of GRV2 and GRV3 
is $10^{-4}-10^{-3}$, showing that the numerical solutions obtained in the present 
study have acceptable accuracy. The explicit values of GRV2 and GRV3 are not 
shown here.
%********************************************
\section{Numerical Results}\label{sec:result}
%******************************************** 
To compute specific models of the magnetized stars, we need specify 
the function forms of the EOS and the function $K$, which determines the distribution of 
the magnetic fields. The polytrope index $n$ is taken to be unity in this study 
because the $n=1$ polytropic EOS reproduces a canonical model of the neutron star. 
As in Ref.~\cite{Shibata:2005ss}, the polytropic constant $K_p$ in cgs units is set to be 
\begin{eqnarray}
K_P = 1.6 \times 10^{5}\,,\label{eq:Kp}
\end{eqnarray}
with which the maximum mass model of the spherical star is characterized by 
$M=1.72M_\odot$ and $R_{\rm cir}=11.8$ km, where $R_{\rm cir}$ denotes 
the circumferential radius of the star at the equator (see Table~ \ref{tab:nonrot-ADMmax}). 
As for the function $K$, we chose two different values of $k$, 
$k=1$ and $k=2$, to investigate how the equilibrium properties depend on the shape of 
the impressed magnetic fields. From equation (\ref{eq:b}), it can be seen that 
the concentration of $|B|$ around its maximum value becomes deeper as the value 
of $k$ increases.

In the numerical scheme in this study, the central density $\rho_c$ and 
the strength of the magnetic field parameter $b$ (see Eq.~(\ref{eq:funcK})) have to be 
given to compute the non-rotating models, i.e., the $\Omega=0$ models. 
For the rotating models, one need specify the axis ratio $r_p/r_e$ in addition to 
$\rho_c$ and $b$, where $r_p$ and $r_e$ are the polar and equatorial radii, 
respectively. To explore properties of the relativistic magnetized star models, 
we construct a large number of the specific models; for the non-rotating cases,  
$26\times301$ models in the parameter space of $(\rho_c,\ b)$, and 
for the rotating cases, $26\times31\times8$ models in the parameter space of 
$(\rho_c,\ b,\ r_p/r_e)$.

\begin{figure*}
  \begin{center}
  \vspace*{40pt}
    \begin{tabular}{c}
      \resizebox{120mm}{!}{\includegraphics{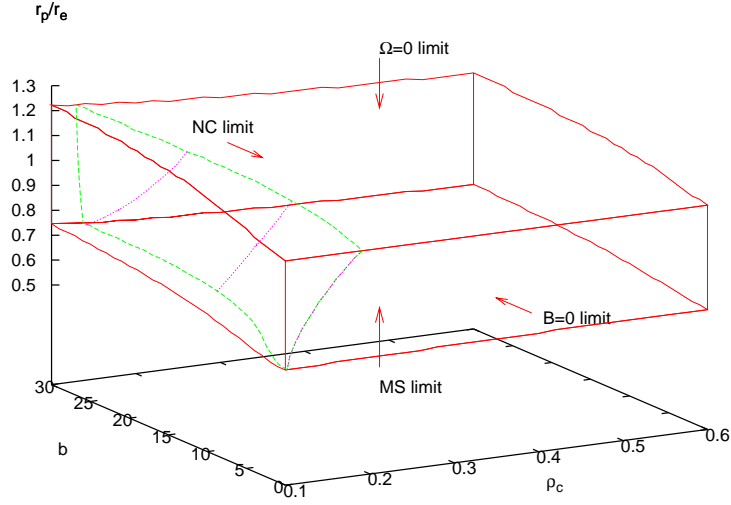}} \\
    \end{tabular}
    \caption{\label{fig:3Dcube}
Solution space for the equilibrium models of the rotating magnetized stars. 
The $x$-, $y$-, and $z$-axis are the central density $\rho_c$, the magnetic 
field strength parameter $b$, and the axis ratio $r_p/r_e$, respectively. A point inside 
the cubic region drawn by the solid lines corresponds to a physically acceptable 
solution computed in this study. 
The sides of the cube labeled by ``$B=0$ limit'' ``$\Omega=0$ limit'' 
``MS limit'' and ``NC limit'' represent the boundaries beyond which there is no 
physically acceptable solution. The surface labeled by ``$B=0$ limit'' corresponds 
to the non-magnetized limit, the surface labeled by ``$\Omega=0$ limit'' to 
the non-rotating limit, and the surface labeled by ``MS limit'' to 
the mass-shedding limit, and the surface labeled by ``NC limit'' to 
the non-convergence limit beyond which any solution cannot converge with the present 
numerical scheme. The surface embedded in the cube that is drawn with the dashed 
line boundary shows a set of equilibrium solutions having the same total baryon rest mass. 
The three dotted curves represent the constant magnetic flux sequences of equilibrium. 
     }
  \end{center}
\end{figure*}

Before describing the numerical results in detail, let us survey the solution space 
for the equilibrium models of the rotating magnetized stars, shown 
in Fig.~\ref{fig:3Dcube}. In this figure, the $x$-, $y$-, and $z$-axis 
are the central density $\rho_c$, the magnetic field strength $b$, 
and the axis ratio $r_p/r_e$, respectively. All these quantities are shown in 
units of $K_p^{n/2}$, as mentioned before. A point inside the cubic region drawn 
by the solid lines corresponds to a physically acceptable solution computed 
in this study. Because we are interested in magnetized neutron stars, equilibrium 
solutions between the two sides of the cube that are given as constant $\rho_c$ 
surfaces are considered. The other four sides of the cube represent the boundaries 
beyond which there is no physically acceptable solution. The $b=0$ surface 
corresponds to the non-magnetized limit, the upper 
surface of nearly $r_p/r_e=$ constant to the non-rotating limit, and the lower  
surface of nearly $r_p/r_e=$ constant to the mass-shedding limit (MS limit). 
Here, the mass-shedding limit means the boundary beyond which no equilibrium 
solution exists because of too strong centrifugal forces, due to which the matter 
is shedding from the equatorial surface of the star. The sixth surface of the cube  
corresponds to the non-convergence limit beyond which any solution cannot 
converge with the present numerical scheme. In other words, for values of $b$ 
larger than some critical value, we fail to obtain a converged solution by 
using our method of solution. The reason of this fail is unclear.

As mentioned before, the constant baryon mass and/or the constant magnetic flux 
sequences of equilibrium stars are important for studying quasi-stationary 
evolution of the isolated neutron star. In Fig.~\ref{fig:3Dcube}, a set of 
equilibrium solutions having the same total baryon rest mass construct 
the surface embedded in the cube that is drawn with the dashed line boundary.  On this 
constant baryon rest mass surface, the three dotted curves represent 
the constant magnetic flux sequences of equilibrium stars. 
%*************************************************
\subsection{Non-rotating models}\label{ssec:nonrot}
\begin{figure*}
  \begin{center}
  \vspace*{40pt}
    \begin{tabular}{cc}
      \resizebox{90mm}{!}{\includegraphics{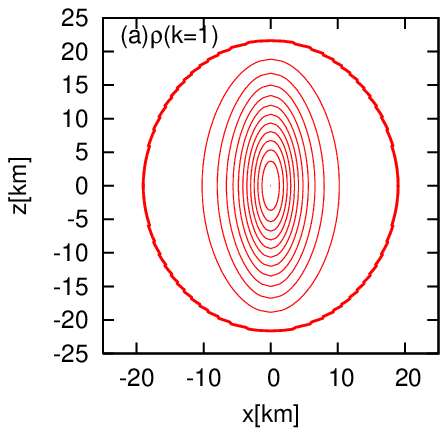}}    &
      \resizebox{90mm}{!}{\includegraphics{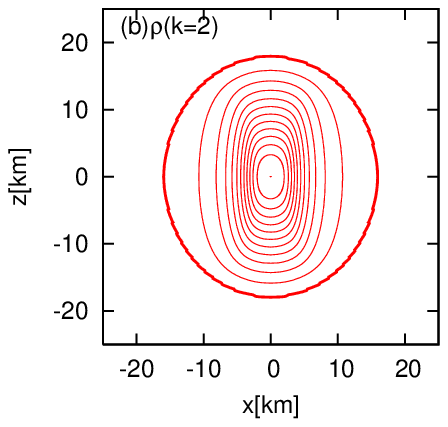}} \\
      \resizebox{90mm}{!}{\includegraphics{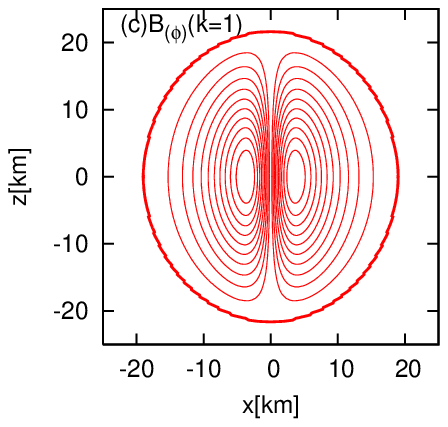}}         &
      \resizebox{90mm}{!}{\includegraphics{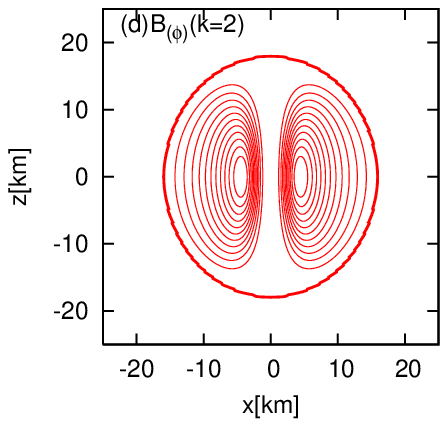}}      \\
    \end{tabular}
    \caption{\label{fig:non-rot}
Distributions of the rest mass density and the magnetic field for the non-rotating 
magnetized stars whose maximum magnetic field strength is $5.0\times10^{17}{\rm G}$. 
(a) Equi-rest-mass-density contours and (c) equi-$B_{(\phi)}$ contours in the meridional 
cross section for the 
$M=1.46M_\odot$, $R_{\rm cir}=20.9{\rm km}$, $H/|W|=0.19$, and $k=1$ model. 
(b) Equi-rest-mass-density contours and (d) equi-$B_{(\phi)}$ contours in the meridional 
cross section for the 
$M=1.51M_\odot$, $R_{\rm cir}=18.0{\rm km}$, $H/|W|=0.15$, and $k=2$ model.
Thick ellipses denote the stellar surface.
}
  \end{center}
\end{figure*}

\begin{figure*}
  \begin{center}
  \vspace*{40pt}
    \begin{tabular}{cc}
      \resizebox{90mm}{!}{\includegraphics{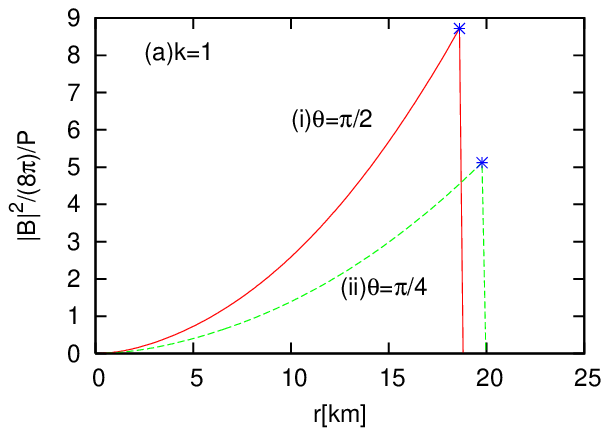}}     &
      \resizebox{90mm}{!}{\includegraphics{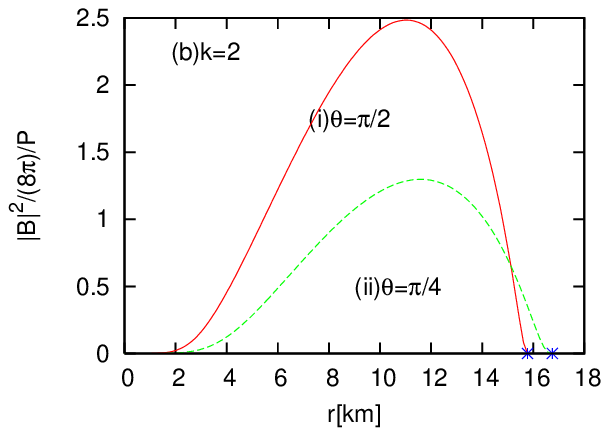}} \\
    \end{tabular}
    \caption{\label{fig:non-rot-Pm-P}
Ratio of the magnetic pressure to the matter pressure $|B|^2/(8\pi P)$ 
at the $\theta=\pi/2$ 
surfaces (solid lines) and the $\theta=\pi/4$ surfaces (dashed lines) 
for the models given in Fig.~\ref{fig:non-rot}, given as functions of $r$. 
(a) $k=1$ and (b) $k=2$. The asterisks indicate the stellar surface. 
}
  \end{center}
\end{figure*}

%*************************************************
First let us consider the static configurations for the 
following two reasons. (1) Since the magnetars and the high field neutron stars 
observed so far are all slow rotators, 
the static models could well approximate to such stars. (2) In the static models, 
one can see purely magnetic effects on the equilibrium properties because there 
is no centrifugal force and all the stellar deformation is attributed 
to the magnetic stress. In Fig.~\ref{fig:non-rot}, we show distributions of 
the rest mass density and the magnetic field in the meridional planes  
for the two static equilibrium stars  
characterized by (1) $\rho_c=6.7\times 10^{14}[{\rm g/cm^3}]$, 
$M=1.46M_\odot$, $R_{\rm cir}=20.9{\rm km}$, $r_p/r_e=1.15$, $\bar{e}=-0.81$,  
$H/|W|=0.19$, and $k=1$, [Panels (a) and (c)] 
and by (2) $\rho_c=6.7\times 10^{14}[{\rm g/cm^3}]$, $M=1.51M_\odot$, 
$R_{\rm cir}=18.0{\rm km}$, $r_p/r_e=1.14$, $\bar{e}=-0.55$, $H/|W|=0.15$, 
and $k=2$ [Panels (b) and (d)]. These two models have the same maximum 
magnetic field strength of $5.0\times10^{17}{\rm G}$. 
In Fig.~\ref{fig:non-rot}, as mentioned, we can confirm that the magnetic stress 
due to the purely toroidal field makes the stellar shape prolate. This is a great 
contrast to the magnetic stress due to the purely poloidal field, for which 
the shape of the star deforms oblately as shown in Ref.~\citep{Chandra:1953}.   
As for the dependence of $k$ on the stellar structures, it is found that 
the density distributions of the stars with $k=1$ are prolately more concentrated 
around the magnetic axis than those of the stars with $k=2$, as can be seen in 
Fig.~\ref{fig:non-rot}. This feature is due to the magnetic pressure 
distributions. As mentioned in the Introduction, the toroidal magnetic field lines 
behave like a rubber belt that is wrapped around the waist of the stars. For 
the $k=1$ case, this tightening due to the toroidal fields is effective 
even near the surface of the star. To see it clearly, in Fig.~\ref{fig:non-rot-Pm-P}, 
we give the ratio of the magnetic pressure to the matter pressure as functions of $r$ 
for the two models shown in Fig.~\ref{fig:non-rot}. In this figure, we see that  
for the $k=1$ case, the magnetic pressure dominates over the matter pressure 
near the stellar surface, while for the  $k=2$ case, the matter pressure 
dominates over the magnetic pressure there.

\begin{figure*}
  \begin{center}
  \vspace*{40pt}
    \begin{tabular}{cc}
      \resizebox{80mm}{!}{\includegraphics{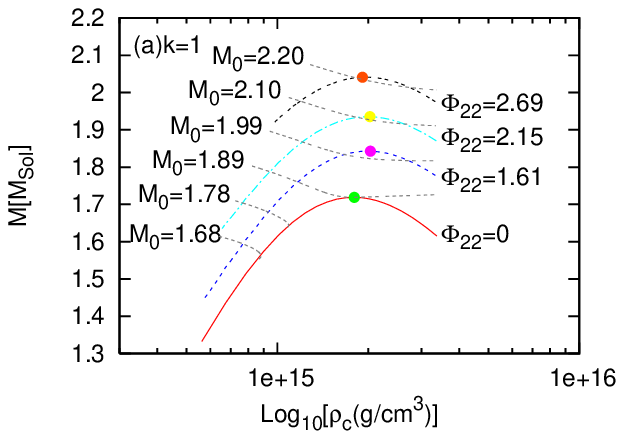}} &
      \resizebox{80mm}{!}{\includegraphics{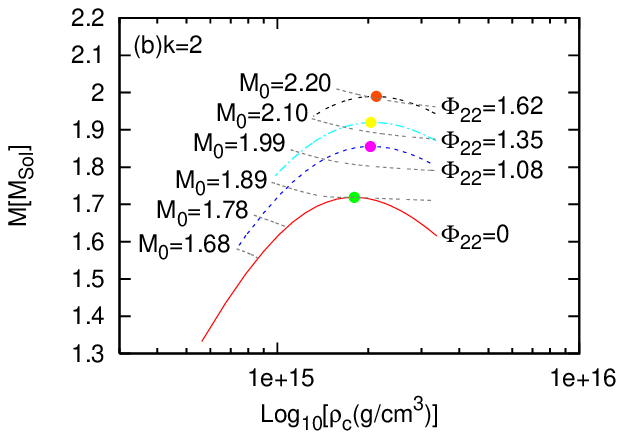}}\\
    \end{tabular}
    \caption{\label{fig:nonrot-qc-ADM}
Central baryon mass density $\rho_c$ versus gravitational mass $M$ along 
the constant magnetic flux equilibrium sequences and the constant magnetic flux 
equilibrium sequences for the non-rotating stars. (a) $k=1$ and (b) $k=2$.
The filled circles indicate the maximum gravitational mass models whose  
global physical quantities are given in Table~\ref{tab:nonrot-ADMmax}. 
Each equilibrium sequence is labeled by its value for $M_0$ or $\Phi_{22}$.
}
  \end{center}
\end{figure*}

In Fig.~\ref{fig:nonrot-qc-ADM}, we show the gravitational mass $M$ of 
static models for several equilibrium sequences of constant $\Phi$ and 
of constant $M_0$ as functions of the central density $\rho_c$. 
Here, $\Phi_{22}$ means the flux normalized by units of $10^{22}$Weber, which 
is a typical value of the magnetic flux for a canonical neutron star model. 
In this figure, the curves labeled by their values of $M_0$ ($\Phi_{22}$)  
indicate the equilibrium sequence along which the values of $M_0$ ($\Phi_{22}$) 
are held constant. The filled circles in Fig.~\ref{fig:nonrot-qc-ADM} 
indicate the maximum mass models of the equilibrium sequences of constant $\Phi$. 
The global physical quantities of these maximum mass models, 
the magnetic flux $\Phi_{22}$, the central baryon mass density $\rho_c$, 
the gravitational mass $M$, the baryon rest mass $M_0$, the circumferential radius 
at the equator $R_{\text{cir}}$, the maximum strength of the magnetic fields 
$B_{\text{max}}$, the ratio of the magnetic energy to the gravitational energy 
$H/|W|$, and the mean deformation rate $\bar{e}$ are summarized in 
Table~\ref{tab:nonrot-ADMmax}. Form this table, we see that 
the strong toroidal magnetic fields of $10^{18}$ G make, roughly $15\%$ increase 
in the maximum gravitational mass of the stars and the highly prolate stellar 
deformation of $\bar{e}\approx -0.7$, even though the magnetic field of 
$10^{18}$ G might be too strong even for a magnetar model. The properties of 
the constant $M_0$ sequences will be discussed in detail later.  

\begin{table*}
\centering
\begin{minipage}{140mm}
\caption{\label{tab:nonrot-ADMmax}
Global physical quantities for the maximum gravitational mass models of the constant 
magnetic flux sequences of the non-rotating stars. 
}
\begin{tabular}{cccccccc}
\hline\hline
$\Phi[10^{22}{\rm Weber}]$          &
$\rho_c[10^{15}{\rm g}/{\rm cm}^3]$ &
$M[M_{\rm \odot}]$     &
$M_0[M_{\rm \odot}]$                &
$R_{\text{cir}}[{\rm km}]$          &
$B_{\text{max}}[10^{18}{\rm G}]$    &
$H/|W|$                             &
$\bar{e}$                          \\
\hline
\multicolumn{8}{c}{k=1}\\
\hline
0.000E+00&1.797E+00&1.719E+00&1.888E+00&1.180E+01&0.000E+00&0.000E+00& 0.000E+00\\
1.616E+00&2.032E+00&1.843E+00&2.014E+00&1.457E+01&1.008E+00&1.253E-01&-4.284E-01\\
2.155E+00&2.026E+00&1.935E+00&2.107E+00&1.667E+01&1.129E+00&1.737E-01&-6.933E-01\\
2.694E+00&1.914E+00&2.041E+00&2.210E+00&1.951E+01&1.168E+00&2.186E-01&-1.012E+00\\
\hline
\multicolumn{8}{c}{k=2}\\
\hline
0.000E+00&1.797E+00&1.719E+00&1.888E+00&1.180E+01&0.000E+00&0.000E+00& 0.000E+00\\
1.077E+00&2.032E+00&1.855E+00&2.055E+00&1.361E+01&8.023E-01&8.068E-02&-3.874E-01\\
1.347E+00&2.039E+00&1.920E+00&2.128E+00&1.444E+01&8.630E-01&1.024E-01&-5.315E-01\\
1.616E+00&2.126E+00&1.990E+00&2.210E+00&1.516E+01&9.205E-01&1.198E-01&-6.721E-01\\
\hline
\end{tabular}
\end{minipage}
\end{table*}

%******************************************
\subsection{Rotating models}\label{ssec:rot}
%******************************************
Next, let us move on to the rotation models. 
For the static stars with the purely toroidal magnetic field, as shown in 
the last subsection, the ratio of the magnetic energy to the gravitational 
energy $H/|W|$ can be of an order of $10^{-1}$, which is larger than 
a typical value of the ratio of the rotation energy to the gravitational energy $T/|W|$ 
for the rotating non-magnetized neutron star models. As shown in \cite{Komatsu:1989}, 
a typical value of $T/|W|$ for the uniformly rotating $n=1$ polytropic neutron star 
models having no magnetic field is of 
an order of $10^{-2}$. This fact implies that a large amount of the magnetic 
energy can be stored in the neutron star with infinite conductivity because 
the mass-shedding like limit does not appear for the magnetic force unlike 
the centrifugal force of the rotating star cases as argued in ~\cite{Cardall:2001}. 
In this paper, as mentioned, we are concerned with equilibrium properties 
of the strongly magnetized stars. Thus, most stellar models treated here are 
magnetic-field-dominated ones in the sense that $H/|W|$ is larger than 
$T/|W|$ even for the mass-shedding models. In such models, 
the effects of the rotation on the stellar structures are necessarily 
supplementary and the basic properties do not highly depend on the rotation parameter.

\begin{figure*}
  \begin{center}
  \vspace*{40pt}
    \begin{tabular}{cc}
      \resizebox{90mm}{!}{\includegraphics{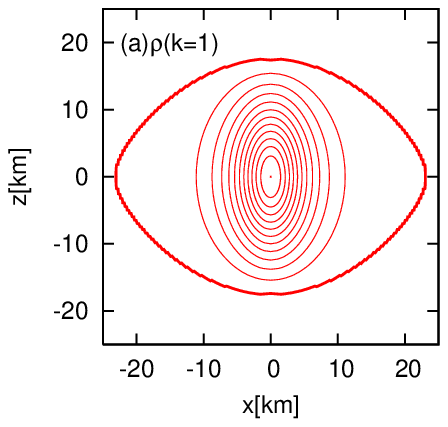}}    &
      \resizebox{90mm}{!}{\includegraphics{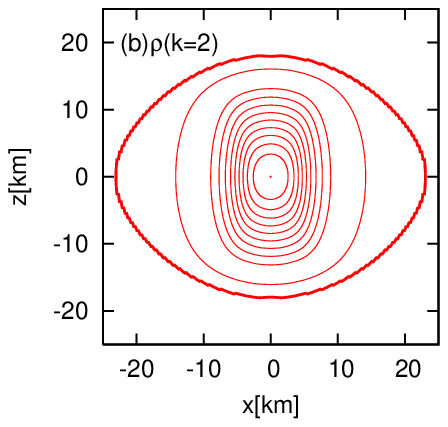}} \\ 
      \resizebox{90mm}{!}{\includegraphics{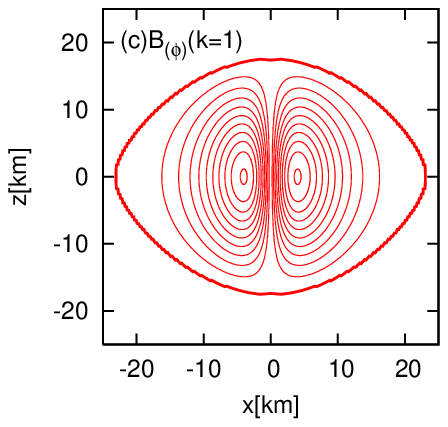}}         &
      \resizebox{90mm}{!}{\includegraphics{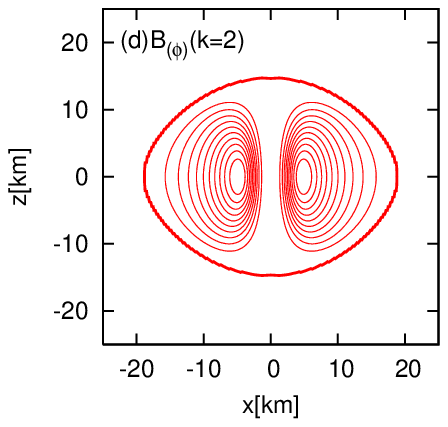}}     \\
    \end{tabular}
    \caption{\label{fig:rot}
Distributions of the rest mass density and the magnetic field for the rotating
magnetized stars whose maximum magnetic field strength is $4.5\times10^{17}{\rm G}$.
(a) Equi-rest-mass-density contours and (c) equi-$B_{(\phi)}$ contours in the meridional
cross section for the 
$M=1.53M_\odot$, $R_{\rm cir}=25.3{\rm km}$, $H/|W|=0.15$, $T/|W|=3.7\times 10^{-2}$
, and $k=1$ model. 
(b) Equi-rest-mass-density contours and (d) equi-$B_{(\phi)}$ contours in the meridional
cross section for the 
$M=1.59M_\odot$, $R_{\rm cir}=21.2{\rm km}$, $H/|W|=0.12$, $T/|W|=5.07\times 10^{-2}$, 
, and $k=2$ model.
Thick ellipses denote the stellar surface.
}
  \end{center}
\end{figure*}

In Fig.~\ref{fig:rot}, we display the typical distributions of 
the rest mass density and the magnetic field for the two rotating equilibrium stars
characterized by (1) $\rho_c=6.7\times 10^{14}[{\rm g/cm^3}]$, 
$M=1.53M_\odot$, $R_{\rm cir}=25.3{\rm km}$, $r_p/r_e=0.75$, $\bar{e}=-0.24$, 
$H/|W|=0.15$, $T/|W|=3.7\times 10^{-2}$, and $k=1$, [Panels (a) and (c)] 
and by (2) $\rho_c=6.7\times 10^{14}[{\rm g/cm^3}]$, 
$M=1.59M_\odot$, $R_{\rm cir}=21.2{\rm km}$, $r_p/r_e=0.78$, $\bar{e}=-6.9\times 10^{-2}$, 
$H/|W|=0.12$, $T/|W|=5.07\times 10^{-2}$, 
and $k=2$ [Panels (b) and (d)]. These  models have the same maximum 
magnetic field strength of $4.5\times10^{17}{\rm G}$. 
Comparing Fig.~\ref{fig:rot} with Fig.~\ref{fig:non-rot}, we see that 
basic properties of the structures of the rotating models are similar to 
those of the static models. Difference between them only appears near 
the equatorial surface of the stars. As shown in Fig.~\ref{fig:rot}, the density 
distributions are stretched from the rotation axis outward due to the centrifugal force 
and the shape of the stellar surface becomes oblate, which can be seen from  
the values of the axis ratio $r_p/r_e$. On the other hand, 
the equi-density contours deep inside the stars are oblate, which can be confirmed 
from the values of the mean deformation rate $\bar{e}$. 
This is because the centrifugal force is the most 
significant agent deforming the stars near the equatorial surface of the star 
since, as shown in equation (\ref{eq:b}), 
the anisotropic magnetic stress of the present models vanishes at the surface of the star. 

%***********************************************************
\subsection{Constant baryon mass sequence}\label{ssec:bcons}
%***********************************************************
In this subsection, we concentrate on the constant baryon mass sequences 
of the magnetized stars. In the static star cases, this condition 
of the constant baryon mass picks out a single sequence of the equilibrium stars. 
Along such an equilibrium sequence, as mentioned, the magnetic flux of the stars cannot 
be held constant. For the rotating star cases, on the other hand, we can select 
a single sequence of the equilibrium stars by keeping the baryon rest mass and 
the magnetic flux constant simultaneously. The equilibrium sequences along which 
the baryon rest mass and the magnetic flux are conserved may model the isolated 
neutron stars that are adiabatically losing angular momentum via the gravitational 
radiation. It should be noted that here, we have omitted evolution of the function 
$K$, which will change its function form during such a process of losing angular 
momentum even though there has been no way to determine it so far. Thus, the constant 
baryon mass 
sequences considered in this study might be oversimplified for investigations of 
the adiabatic evolution of the isolated neutron stars. However, the inclusion of 
the evolution effects of the function $K$ is beyond the scope of the present study. 

First we consider the constant baryon mass sequences of the non-rotating stars 
to clearly understand the basic properties of the simplest equilibrium 
sequences. Then we move on to the rotating cases. As done in Ref.~\cite{Cook:1992}, 
we divide the equilibrium sequences of the magnetized stars into two classes, 
{\it normal} and {\it supramassive} equilibrium sequences. In this study, 
the {\it normal} ({\it supramassive}) equilibrium sequence is defined as 
an equilibrium sequence whose baryon rest mass is smaller (larger) than the maximum 
baryon rest mass of the non-magnetized and non-rotating stars. As shown in 
Fig.~\ref{fig:nonrot-qc-ADM}, the maximum baryon rest mass of the non-rotating and 
non-magnetized stars for the present EOS is given by 
$M_0=1.89M_\odot$. Therefore, the normal (supramassive) equilibrium sequences 
are simply defined as equilibrium sequences with $M_0 < 1.89M_\odot$ 
($M_0 > 1.89M_\odot$).

%********************************************
\subsubsection{Static cases}
%********************************************
In Fig.~\ref{fig:nonrot-qc-ADM}, one can see how the constant 
baryon mass sequences of the non-rotating stars are drawn on the $\rho_c$--$M$ plane. 
In this figure, the two constant baryon mass sequences characterized 
by $M_0=1.68M_\odot$ and $1.78M_\odot$ refer to the normal equilibrium sequences, while 
the three constant baryon mass sequences characterized by $M_0=1.99M_\odot$, 
$2.10M_\odot$, and $2.20M_\odot$ refer to the supramassive equilibrium sequences.  
In Fig.~\ref{fig:nonrot-qc-ADM}, one can confirm that the two normal equilibrium 
sequences of $M_0=1.68M_\odot$ and $1.78M_\odot$ indeed end at the equilibrium stars  
having $\Phi=0$ (no magnetic field) and that the supramassive equilibrium sequences never 
intersect with the curves given by $\Phi=0$. Note that all the end points of the 
equilibrium sequences of the non-rotating magnetized stars correspond to the ``NC limit'' 
in Fig.~\ref{fig:3Dcube}.

\begin{figure*}
  \begin{center}
  \vspace*{40pt}
    \begin{tabular}{cc}
      \resizebox{75mm}{!}{\includegraphics{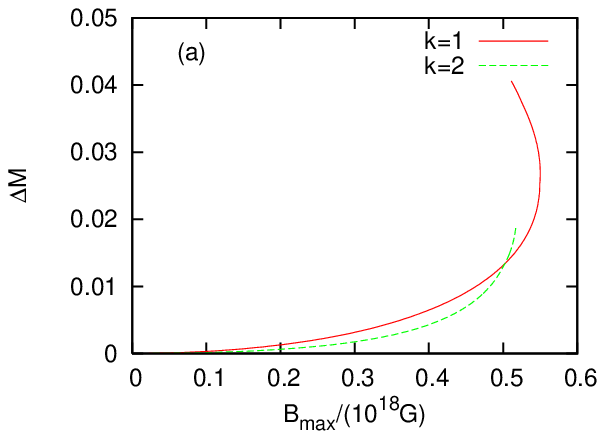}} &
      \resizebox{75mm}{!}{\includegraphics{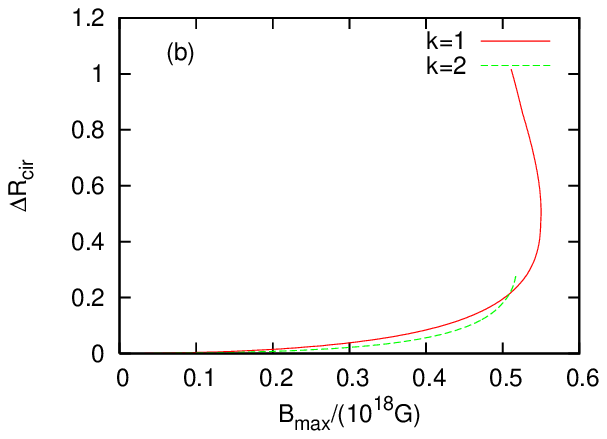}}   \\
      \resizebox{75mm}{!}{\includegraphics{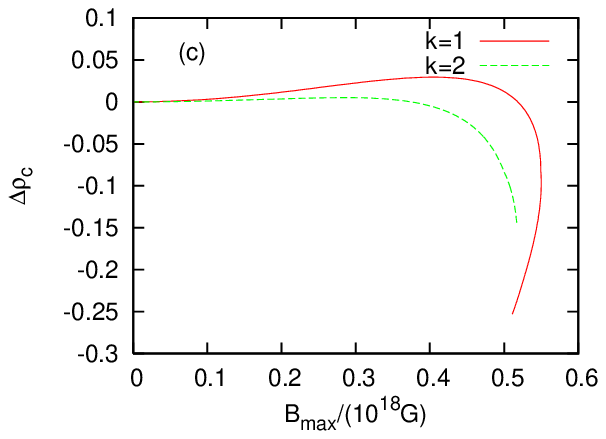}} &
      \resizebox{75mm}{!}{\includegraphics{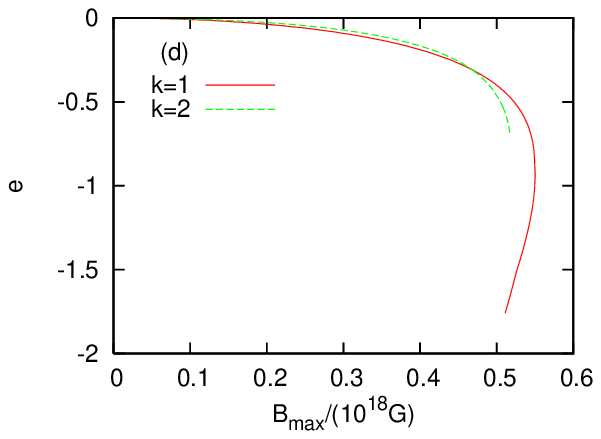}}   \\
    \end{tabular}
    \caption{\label{fig:Mb-const-nom}
Global physical quantities (a) the change rate of the gravitational mass 
$\Delta M$, (b) the change rate of the circumferential radius at the equator 
$\Delta R_{\rm cir}$, (c) the change rate of the central rest mass density 
$\Delta\rho_c$, and (d) the mean deformation rate $\bar{e}$ 
along the normal equilibrium sequences of the non-rotating stars characterized by 
$M_0=1.68 M_{\odot}$, given as functions of $B_{\rm max}$. 
    }
  \end{center}
\end{figure*}

\begin{figure*}
  \begin{center}
  \vspace*{40pt}
    \begin{tabular}{cc}
      \resizebox{75mm}{!}{\includegraphics{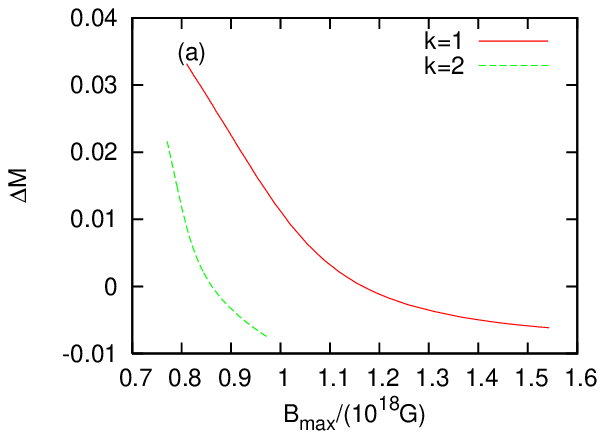}} &
      \resizebox{75mm}{!}{\includegraphics{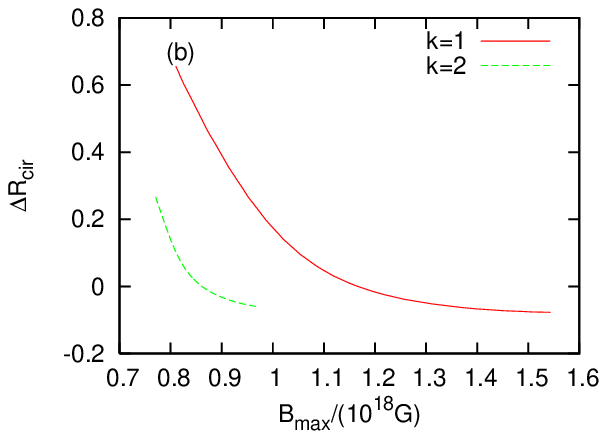}}   \\
      \resizebox{75mm}{!}{\includegraphics{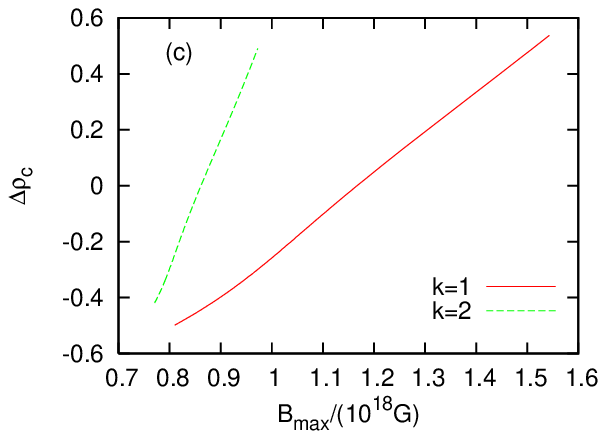}} &
      \resizebox{75mm}{!}{\includegraphics{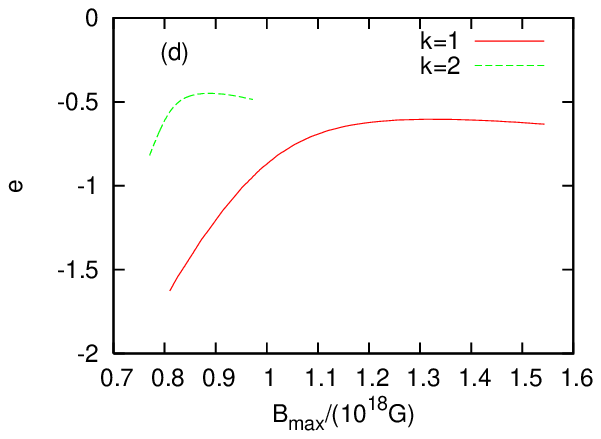}}   \\
    \end{tabular}
    \caption{\label{fig:Mb-const-sup}
     Same as Fig.~\ref{fig:Mb-const-nom}, but for the supramassive 
sequence with $M_0=2.10M_\odot$.
    }
  \end{center}
\end{figure*}

In Figs.~\ref{fig:Mb-const-nom} and \ref{fig:Mb-const-sup}, the global physical 
quantities, $\Delta M$, $\Delta R_{cir}$, $\Delta \rho_c$, and $\bar{e}$ are given as 
functions of the maximum values of $|B|$ inside the star, $B_{\rm max}$, for the normal 
equilibrium sequences with $M_0=1.68 M_\odot$ and for the supramassive equilibrium 
sequences with $M_0=2.10 M_\odot$, respectively. In these figures, 
\begin{equation}
\Delta Q(B_{\rm max},M_0) \equiv 
{Q(B_{\rm max},M_0)-Q(|B|=0,M_0) \over Q(|B|=0, M_0)}\,, 
\end{equation}
where $Q$ stands for a physical quantity for the constant baryon mass sequences 
of the non-rotating stars. Numerical values of the global physical quantities,  
the magnetic flux $\Phi$, the central rest mass density $\rho_c$, the gravitational 
mass $M$, the circumferential radius $R_{\text{cir}}$, the maximum strength of the 
magnetic field $B_{\text{max}}$, the ratio of the magnetic energy to the 
gravitational energy $H/|W|$, and the mean deformation rate $\bar{e}$, 
for the selected models,
which are displayed in Figs.~\ref{fig:Mb-const-nom} and \ref{fig:Mb-const-sup},  
are also summarized in Tables~\ref{tab:Mb-const-nom} and ~\ref{tab:Mb-const-sup}. 
From Figs.~\ref{fig:Mb-const-nom} and \ref{fig:Mb-const-sup}, we see that, qualitatively, 
basic behavior of $\Delta M$, $\Delta R_{cir}$, $\Delta \rho_c$, $\bar{e}$ does not 
highly depend on the value of $k$. Comparing Fig.~\ref{fig:Mb-const-nom} with 
Fig.~\ref{fig:Mb-const-sup}, we find that as the value of $B_{\rm max}$ is varied,  
the global changes in the global physical quantities for the normal sequences are 
of opposite directions to those for the supramassive equilibrium sequences.

Before going to the the rotating cases, in which we consider the equilibrium sequences along 
which the baryon rest mass and the magnetic flux are held constants, 
it is useful to investigate non-rotating solutions referred to these equilibrium 
sequences of the constant baryon rest mass and magnetic flux. Plots given in 
Fig.~\ref{fig:nonrot-qc-ADM} tell us the number of such solutions. If 
the values of the baryon rest mass and magnetic flux are specified, one can 
pick out two curves corresponding to the constant baryon rest mass and the constant 
magnetic flux. Then, the intersections of these two curves indicate 
the members of the constant baryon rest mass and magnetic flux sequences in 
the non-rotating limit. 
For the solutions of the normal equilibrium sequences, 
characterized by $M_0 < 1.89$, it can be seen that there is a single intersection 
of the two curves. 
(Note that we only consider the solutions in the lower central density region, 
even though there are other solutions in the higher central density region.) 
For example, in the left panel of Fig.~\ref{fig:nonrot-qc-ADM}, the $M_0=1.78$ 
curve intersects once to the $\Phi_{22}=1.61$ curve. For the solutions of the supramassive 
equilibrium sequences, characterized by $M_0 > 1.89$, on the other hand, we see 
that there are the following three possibilities. (1) There is no intersection between 
the two curves, showing that the non-rotating limit does not exist. (2) There is a 
single intersection between the two curves, which is only possible when one 
curve is tangent to the other. In this case, a unique limit of no rotation exists. 
(3) There are two intersections between the two curves, showing that there are two 
different limits of no rotation in this case. For example, in the left panel of 
Fig.~\ref{fig:nonrot-qc-ADM}, let us consider the constant baryon rest mass sequence 
of $M_0=2.10$. If we chose $\Phi_{22}$ to be $2.15$, $2.05$, and $1.61$, then  
the numbers of the intersection of the two curves are, respectively, two, one, and 
zero, as can be seen from Fig.~\ref{fig:nonrot-qc-ADM}. Thus, it is found that 
the supramassive equilibrium sequences with $(M_0,\ \Phi_{22})=(2.10,\ 1.61) $, 
$(2.10,\ 2.05)$, and $(2.10,\ 2.15)$ have no non-rotation limit, the single non-rotating limit, 
and double non-rotation limits, respectively. One important consequence of this fact  
is that there are two disconnected branches of the equilibrium sequences characterized by 
the same baryon rest mass and the magnetic flux for some supramassive equilibrium 
sequences, as discussed in the next subsection.

\begin{table*}
\centering
\begin{minipage}{140mm}
\caption{\label{tab:Mb-const-nom}
Global physical quantities for the normal equilibrium sequences of the non-rotating 
stars with $M_0=1.68M_\odot$. }
\begin{tabular}{ccccccc}
\hline\hline
$\Phi[10^{22}{\rm Weber}]$          &
$\rho_c[10^{15}{\rm g}/{\rm cm}^3]$ &
$M[M_{\rm \odot}]$     &
$R_{\text{cir}}[{\rm km}]$          &
$B_{\text{max}}[10^{18}{\rm G}]$    &
$H/|W|$                             &
$\bar{e}$                          \\
\hline
\multicolumn{7}{c}{$k=1$}\\
\hline
0.000E+00&8.569E-01&1.551E+00&1.430E+01&0.000E+00&0.000E+00& 0.000E+00\\
1.171E+00&8.538E-01&1.576E+00&1.786E+01&5.222E-01&1.307E-01&-4.975E-01\\
1.754E+00&7.371E-01&1.600E+00&2.340E+01&5.458E-01&2.225E-01&-1.151E+00\\
2.120E+00&6.400E-01&1.614E+00&2.885E+01&5.111E-01&2.772E-01&-1.758E+00\\
\hline
\multicolumn{7}{c}{$k=2$}\\
\hline
0.000E+00&8.569E-01&1.551E+00&1.430E+01&0.000E+00&0.000E+00& 0.000E+00\\
7.239E-01&8.178E-01&1.566E+00&1.611E+01&4.761E-01&8.635E-02&-3.372E-01\\
9.424E-01&7.677E-01&1.575E+00&1.735E+01&5.089E-01&1.232E-01&-5.391E-01\\
1.074E+00&7.335E-01&1.580E+00&1.825E+01&5.170E-01&1.454E-01&-6.831E-01\\
\hline
\end{tabular}
\end{minipage}
\end{table*}

\begin{table*}
\centering
\begin{minipage}{140mm}
\caption{\label{tab:Mb-const-sup}
Global physical quantities for the supramassive equilibrium sequences of 
the non-rotating stars with $M_0=2.10M_\odot$.}
\begin{tabular}{ccccccc}
\hline\hline
$\Phi[10^{22}{\rm Weber}]$          &
$\rho_c[10^{15}{\rm g}/{\rm cm}^3]$ &
$M[M_{\rm \odot}]$     &
$R_{\text{cir}}[{\rm km}]$          &
$B_{\text{max}}[10^{18}{\rm G}]$    &
$H/|W|$                             &
$\bar{e}$                          \\
\hline
\multicolumn{7}{c}{$k=1$}\\
\hline
2.178E+00&1.779E+00&1.936E+00&1.765E+01&1.048E+00&1.842E-01&-7.650E-01\\
2.430E+00&1.415E+00&1.958E+00&2.083E+01&9.368E-01&2.230E-01&-1.064E+00\\
2.650E+00&1.222E+00&1.974E+00&2.374E+01&8.659E-01&2.532E-01&-1.351E+00\\
2.838E+00&1.092E+00&1.987E+00&2.651E+01&8.102E-01&2.778E-01&-1.626E+00\\
\hline
\multicolumn{7}{c}{$k=2$}\\
\hline
1.250E+00&1.980E+00&1.896E+00&1.424E+01&8.360E-01&9.620E-02&-4.833E-01\\
1.365E+00&1.625E+00&1.909E+00&1.542E+01&8.044E-01&1.162E-01&-5.874E-01\\
1.460E+00&1.468E+00&1.918E+00&1.622E+01&7.897E-01&1.307E-01&-6.753E-01\\
1.538E+00&1.371E+00&1.925E+00&1.685E+01&7.793E-01&1.421E-01&-7.510E-01\\
\hline
\end{tabular}
\end{minipage}
\end{table*}

%***********************************************
\subsubsection{Rotating cases}
%***********************************************

\begin{figure*}
  \begin{center}
  \vspace*{40pt}
    \begin{tabular}{cc}
      \resizebox{75mm}{!}{\includegraphics{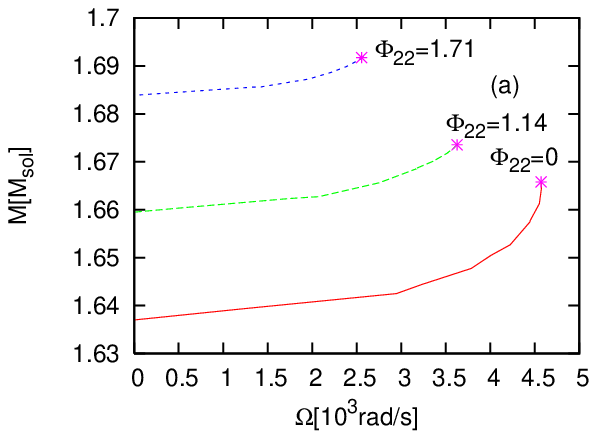}} &
      \resizebox{75mm}{!}{\includegraphics{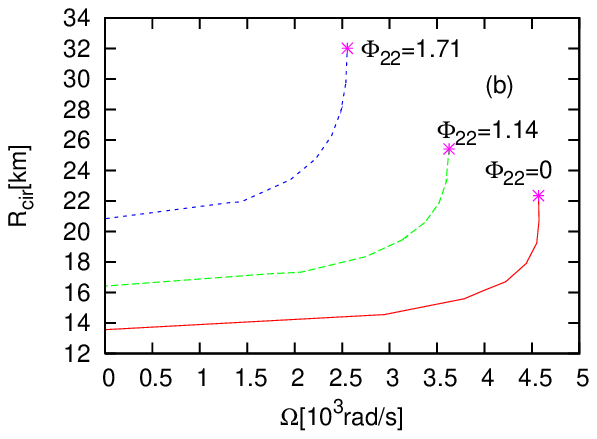}}   \\
      \resizebox{75mm}{!}{\includegraphics{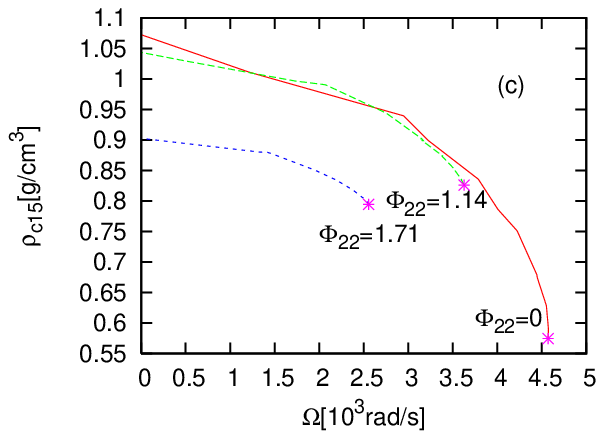}} &
      \resizebox{75mm}{!}{\includegraphics{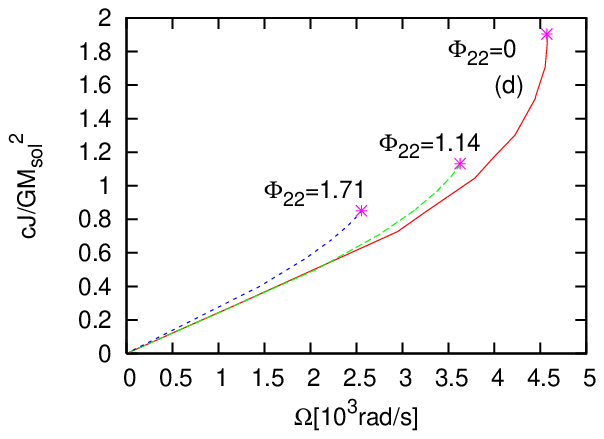}}   \\
      \resizebox{75mm}{!}{\includegraphics{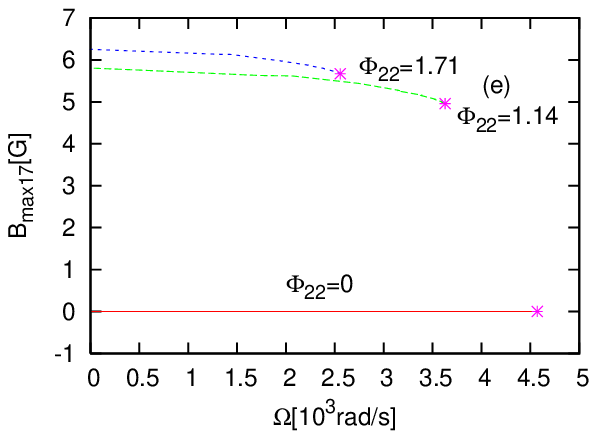}} &
      \resizebox{75mm}{!}{\includegraphics{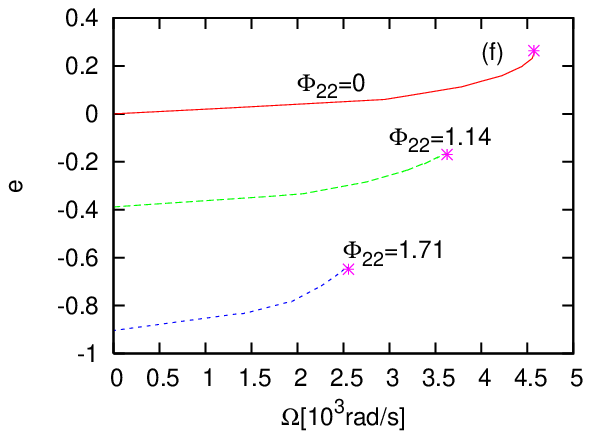}}   \\
    \end{tabular}
    \caption{\label{fig:Mb-Phi-const-nom}
Global physical quantities $M$, $R_{cir}$, $\rho_c$, $J$, $B_{\rm max}$,
and $\bar{e}$ for the constant baryon mass and magnetic flux equilibrium 
sequences with $k=1$, given as functions of $\Omega$. All the equilibrium 
sequences are referred to the normal equilibrium sequences characterized
by the constant baryon rest mass $M_0=1.78M_\odot$. Each sequence is 
labeled by its value of $\Phi_{22}$. The asterisks indicate 
the mass-shedding models. 
    }
  \end{center}
\end{figure*}

\begin{figure*}
  \begin{center}
  \vspace*{40pt}
    \begin{tabular}{cc}
      \resizebox{75mm}{!}{\includegraphics{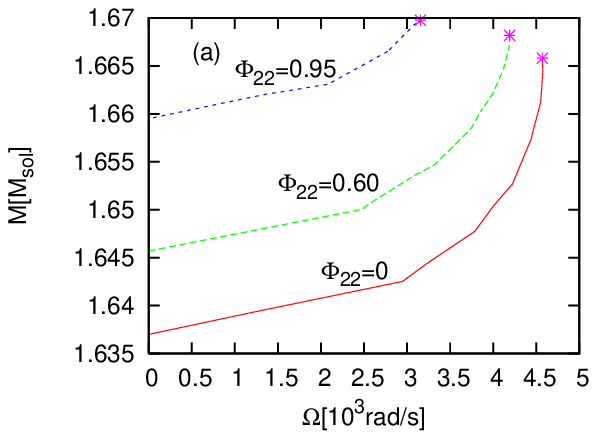}} &
      \resizebox{75mm}{!}{\includegraphics{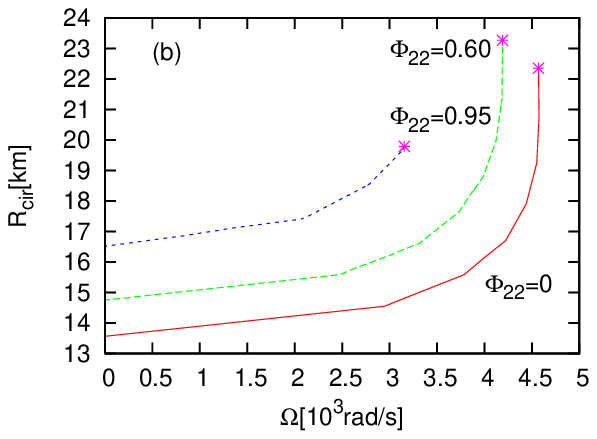}}   \\
      \resizebox{75mm}{!}{\includegraphics{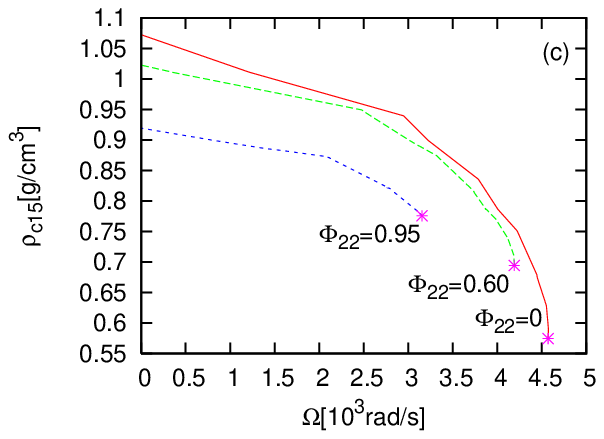}} &
      \resizebox{75mm}{!}{\includegraphics{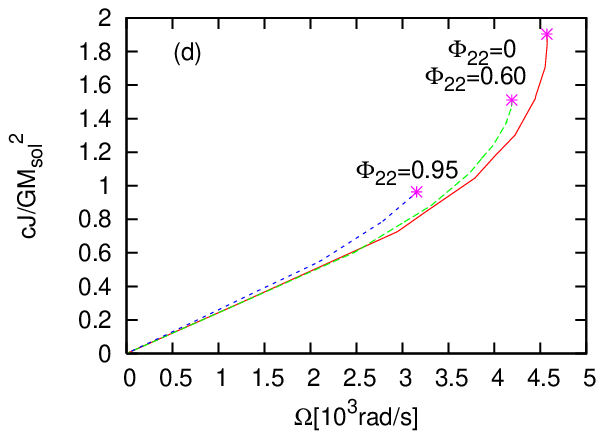}}   \\
      \resizebox{75mm}{!}{\includegraphics{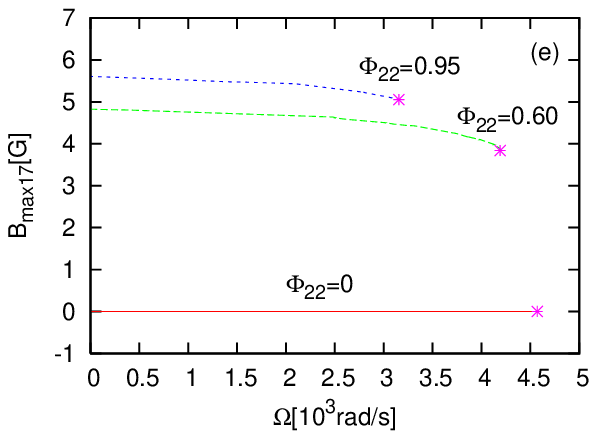}} &
      \resizebox{75mm}{!}{\includegraphics{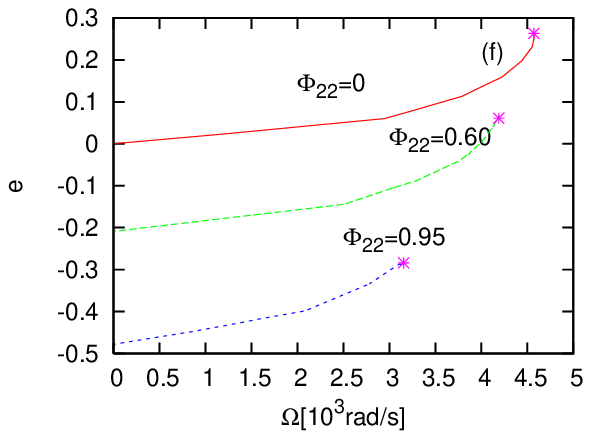}}   \\
    \end{tabular}
    \caption{\label{fig:Mb-Phi-const-nom-k2}
Same as Fig.~\ref{fig:Mb-Phi-const-nom}, but for $k=2$.
    }
  \end{center}
\end{figure*}

First, let us consider the normal equilibrium sequences of the magnetized 
rotating stars. 
In Figs. \ref{fig:Mb-Phi-const-nom} and \ref{fig:Mb-Phi-const-nom-k2}, 
we show the global physical quantities $M$, $R_{cir}$, $\rho_c$, $J$, $B_{\rm max}$, 
and $\bar{e}$ as functions of $\Omega$ for the constant baryon mass and magnetic flux 
equilibrium sequences with $k=1$ and $k=2$, respectively. All the equilibrium 
sequences given in these figures are of the normal equilibrium sequences characterized 
by the constant baryon rest mass given by $M_0=1.78M_\odot$. In Figs. \ref{fig:Mb-Phi-const-nom} 
and \ref{fig:Mb-Phi-const-nom-k2}, each curve is labeled by its value of $\Phi_{22}$ 
which is held constant along the equilibrium sequence. For the sake of comparison, 
the physical quantities for the non-magnetized equilibrium sequences ($\Phi_{22}=0$) 
are also given in Figs. \ref{fig:Mb-Phi-const-nom} and \ref{fig:Mb-Phi-const-nom-k2}. 
Note that the constant baryon rest mass equilibrium sequences of the non-magnetized 
polytropic  stars have been argued in detail in Ref.~\cite{Cook:1992}. 
Numerical values of the global physical quantities, the central rest mass density 
$\rho_c$, the gravitational mass $M$, the circumferential radius $R_{\text{cir}}$,
the angular velocity $\Omega$, the angular momentum $J$, the ratio of the rotation 
energy to the gravitational energy $T/|W|$, the ratio of the magnetic energy to 
the gravitational energy $H/|W|$, the mean deformation rate $\bar{e}$, and 
the maximum strength of the magnetic field $B_{\text{max}}$, 
for some selected models are 
tabulated in Tables \ref{tab:Mb-const-nom-rot} and \ref{tab:Mb-const-nom-rot-k2} 
for the cases of $k=1$ and $k=2$, respectively.

In Figs. \ref{fig:Mb-Phi-const-nom} and \ref{fig:Mb-Phi-const-nom-k2}, 
as discussed, we see that all the normal equilibrium sequences begin at the non-rotating 
equilibrium stars and continue to the mass-shedding limits, at which 
the sequences end, as the angular velocity increases. (In Figs. 
\ref{fig:Mb-Phi-const-nom} and \ref{fig:Mb-Phi-const-nom-k2}, the asterisks indicate 
the mass-shedding models.) It is also found from these figures that the behavior  
of the normal equilibrium sequences as functions of $\Omega$ are basically independent of 
the values of $\Phi_{22}$ and $k$. As the angular velocity is increased, 
the gravitational masses and the circumferential radii increase, while the central 
densities and the maximum strength of the magnetic fields decrease because of 
the baryon rest mass constancy and of the magnetic flux constancy, respectively. 
The angular momenta are monotonically increasing functions of the angular velocity, 
which implies that the angular momentum loss via gravitational radiation results 
in the spin down of the star for the normal equilibrium sequences. 

\begin{table*}
\centering
\begin{minipage}{140mm}
\caption{\label{tab:Mb-const-nom-rot}
Global physical quantities for the normal equilibrium sequences of 
the rotating stars with $M_0=1.78M_\odot$ and $k=1$. 
}
\begin{tabular}{ccccccccc}
\hline\hline
$\rho_c[10^{15}{\rm g}/{\rm cm}^3]$ &
$M[M_{\rm \odot}]$     &
$R_{\text{cir}}[{\rm km}]$          &
$\Omega[10^3{\rm rad/s}]$           &
$cJ/G M_\odot^2$                    &
$T/|W|$                             &
$H/|W|$                             &
$\bar{e}$                           &
$B_{\text{max}}[10^{17}{\rm G}]$    \\
\hline\hline
\multicolumn{9}{c}{$\Phi_{22}=0$}\\
\hline
5.745E-01&1.666E+00&2.235E+01&4.571E+00&1.905E+00&9.657E-02&0.000E+00& 2.636E-01&0.000E+00\\
6.286E-01&1.661E+00&1.925E+01&4.552E+00&1.707E+00&8.224E-02&0.000E+00& 2.310E-01&0.000E+00\\
7.864E-01&1.650E+00&1.616E+01&4.008E+00&1.179E+00&4.497E-02&0.000E+00& 1.368E-01&0.000E+00\\
8.987E-01&1.644E+00&1.490E+01&3.230E+00&8.355E-01&2.444E-02&0.000E+00& 7.764E-02&0.000E+00\\
1.073E+00&1.637E+00&1.356E+01&0.000E+00&0.000E+00&0.000E+00&0.000E+00& 1.928E-04&0.000E+00\\
\hline
\multicolumn{9}{c}{$\Phi_{22}=1.14$}\\
\hline
8.261E-01&1.674E+00&2.542E+01&3.627E+00&1.131E+00&4.242E-02&1.194E-01&-1.697E-01&4.957E+00\\
8.739E-01&1.670E+00&2.065E+01&3.386E+00&9.709E-01&3.279E-02&1.180E-01&-2.053E-01&5.162E+00\\
9.017E-01&1.668E+00&1.952E+01&3.154E+00&8.631E-01&2.663E-02&1.170E-01&-2.395E-01&5.275E+00\\
9.913E-01&1.663E+00&1.732E+01&2.035E+00&4.937E-01&9.463E-03&1.133E-01&-3.339E-01&5.621E+00\\
1.044E+00&1.660E+00&1.642E+01&0.000E+00&0.000E+00&0.000E+00&1.111E-01&-3.883E-01&5.809E+00\\
\hline
\multicolumn{9}{c}{$\Phi_{22}=1.71$}\\
\hline
7.947E-01&1.692E+00&3.202E+01&2.554E+00&8.514E-01&2.516E-02&2.001E-01&-6.485E-01&5.671E+00\\
8.314E-01&1.689E+00&2.482E+01&2.227E+00&6.776E-01&1.690E-02&1.988E-01&-7.263E-01&5.871E+00\\
8.504E-01&1.687E+00&2.340E+01&1.957E+00&5.703E-01&1.230E-02&1.979E-01&-7.791E-01&5.977E+00\\
8.782E-01&1.686E+00&2.200E+01&1.453E+00&4.019E-01&6.310E-03&1.973E-01&-8.298E-01&6.125E+00\\
9.035E-01&1.684E+00&2.084E+01&0.000E+00&0.000E+00&0.000E+00&1.959E-01&-9.042E-01&6.258E+00\\
\hline
\end{tabular}
\end{minipage}
\end{table*}

\begin{table*}
\centering
\begin{minipage}{140mm}
\caption{\label{tab:Mb-const-nom-rot-k2}
Global physical quantities for the normal equilibrium sequences of
the rotating stars with $M_0=1.78M_\odot$ and $k=2$.
}
\begin{tabular}{ccccccccc}
\hline\hline
$\rho_c[10^{15}{\rm g}/{\rm cm}^3]$ &
$M[M_{\rm \odot}]$     &
$R_{\text{cir}}[{\rm km}]$          &
$\Omega[10^3{\rm rad/s}]$           &
$cJ/G M_\odot^2$                    &
$T/|W|$                             &
$H/|W|$                             &
$\bar{e}$                           &
$B_{\text{max}}[10^{17}{\rm G}]$    \\
\hline\hline
\multicolumn{9}{c}{$\Phi_{22}=0.60$}\\
\hline
6.943E-01&1.668E+00&2.327E+01&4.191E+00&1.511E+00&6.772E-02&6.524E-02& 6.091E-02&3.842E+00\\
7.115E-01&1.667E+00&2.133E+01&4.185E+00&1.464E+00&6.459E-02&6.483E-02& 6.430E-02&3.898E+00\\
8.137E-01&1.658E+00&1.769E+01&3.742E+00&1.083E+00&3.949E-02&6.203E-02&-4.386E-02&4.245E+00\\
9.465E-01&1.650E+00&1.562E+01&2.503E+00&6.106E-01&1.373E-02&5.869E-02&-1.449E-01&4.617E+00\\
1.024E+00&1.646E+00&1.475E+01&0.000E+00&0.000E+00&0.000E+00&5.666E-02&-2.079E-01&4.837E+00\\
\hline
\multicolumn{9}{c}{$\Phi_{22}=0.95$}\\
\hline
7.757E-01&1.670E+00&1.979E+01&3.156E+00&9.639E-01&3.195E-02&1.165E-01&-2.838E-01&5.052E+00\\
8.181E-01&1.667E+00&1.863E+01&2.811E+00&8.000E-01&2.291E-02&1.150E-01&-3.297E-01&5.222E+00\\
8.710E-01&1.663E+00&1.745E+01&2.112E+00&5.532E-01&1.146E-02&1.130E-01&-3.955E-01&5.428E+00\\
9.001E-01&1.661E+00&1.684E+01&7.827E-01&2.024E-01&4.036E-03&1.118E-01&-4.507E-01&5.541E+00\\
9.198E-01&1.659E+00&1.651E+01&0.000E+00&0.000E+00&0.000E+00&1.109E-01&-4.785E-01&5.607E+00\\
\hline
\end{tabular}
\end{minipage}
\end{table*}

\begin{figure*}
  \begin{center}
  \vspace*{40pt}
    \begin{tabular}{cc}
      \resizebox{75mm}{!}{\includegraphics{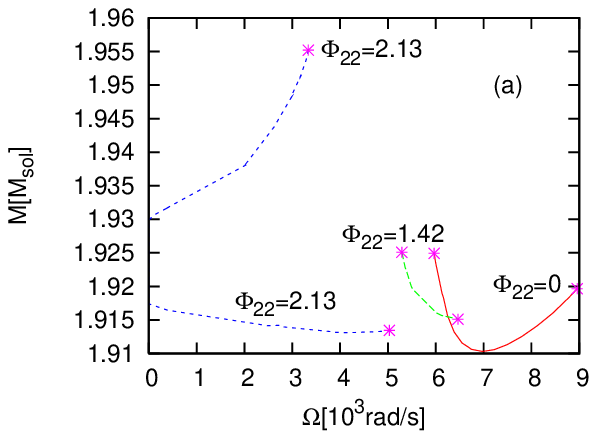}} &
      \resizebox{75mm}{!}{\includegraphics{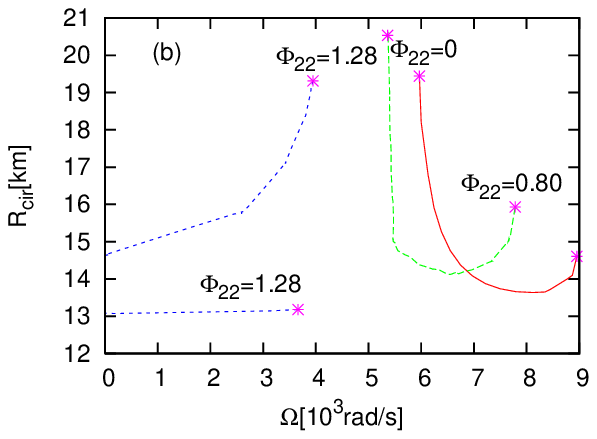}}   \\
      \resizebox{75mm}{!}{\includegraphics{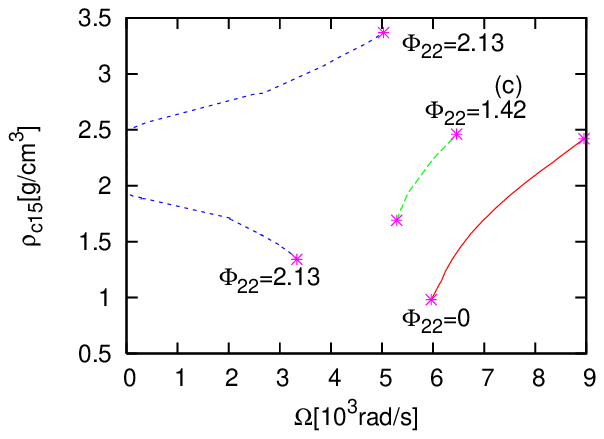}} &
      \resizebox{75mm}{!}{\includegraphics{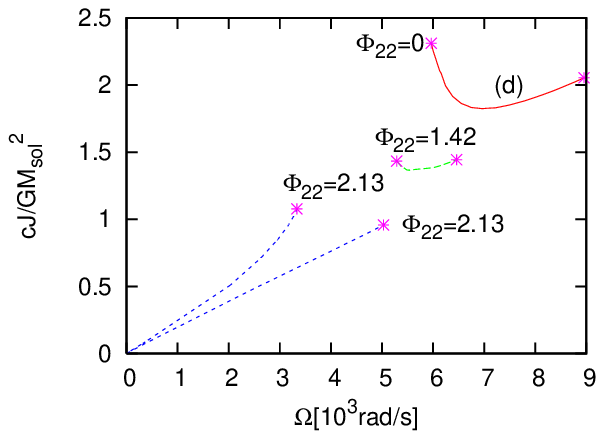}}   \\
      \resizebox{75mm}{!}{\includegraphics{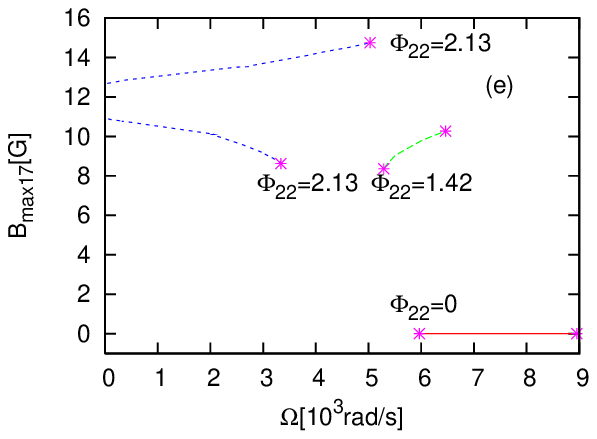}}   &
      \resizebox{75mm}{!}{\includegraphics{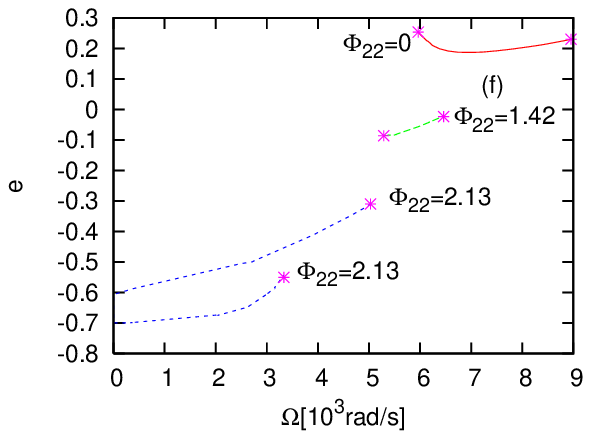}}   \\
    \end{tabular}
    \caption{\label{fig:Mb-Phi-const-sup}
Global physical quantities $M$, $R_{cir}$, $\rho_c$, $J$, $B_{\rm max}$,
and $\bar{e}$ for the constant baryon mass and magnetic flux equilibrium 
sequences with $k=1$, given as functions of $\Omega$.  All 
the equilibrium sequences are referred to the supramassive equilibrium 
sequences characterized by the constant baryon rest mass $M_0=2.10M_\odot$. 
Each sequence is labeled by its value of $\Phi_{22}$. 
The asterisks indicate the mass-shedding models. 
    }
  \end{center}
\end{figure*}

\begin{figure*}
  \begin{center}
  \vspace*{40pt}
    \begin{tabular}{cc}
      \resizebox{75mm}{!}{\includegraphics{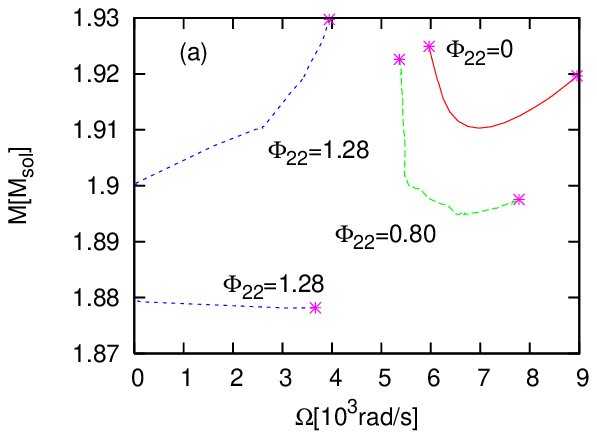}} &
      \resizebox{75mm}{!}{\includegraphics{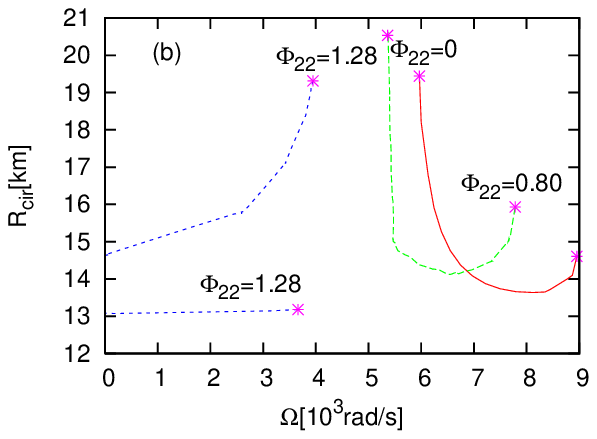}}   \\
      \resizebox{75mm}{!}{\includegraphics{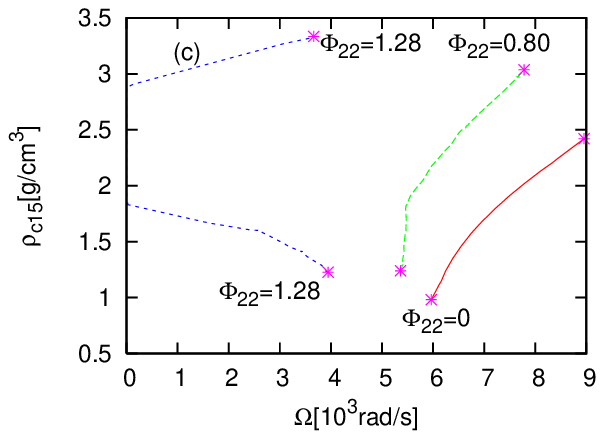}} &
      \resizebox{75mm}{!}{\includegraphics{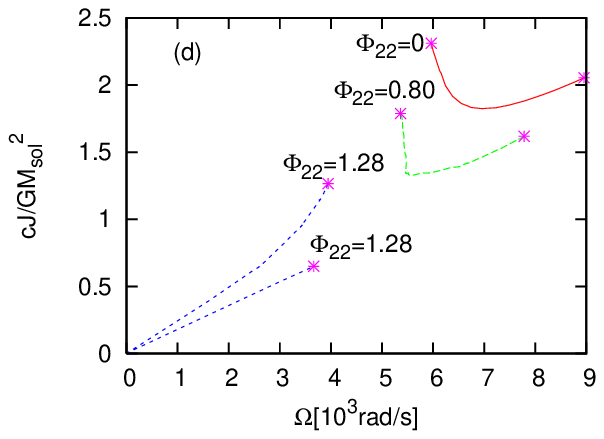}}   \\
      \resizebox{75mm}{!}{\includegraphics{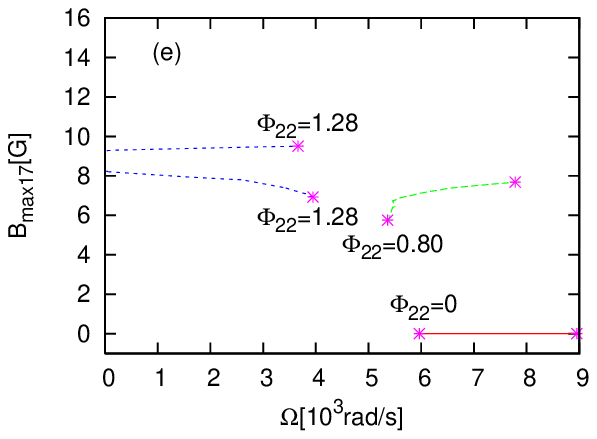}}   &
      \resizebox{75mm}{!}{\includegraphics{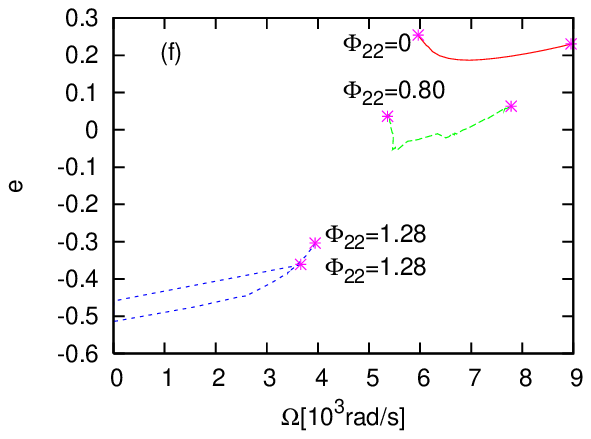}}   \\
    \end{tabular}
    \caption{\label{fig:Mb-Phi-const-sup-k2}
    Same as Fig.~\ref{fig:Mb-Phi-const-sup}, but for $k=2$.
    }
  \end{center}
\end{figure*}

Now consider the supramassive equilibrium sequences of the magnetized rotating stars.
Figures \ref{fig:Mb-Phi-const-sup} and \ref{fig:Mb-Phi-const-sup-k2} give the plots 
of the global physical quantities $M$, $R_{\text{cir}}$, $\rho_c$, $J$, $B_{\rm max}$,
and $\bar{e}$ as functions of $\Omega$ for the constant baryon mass and magnetic flux
equilibrium sequences with $k=1$ and $k=2$, respectively. All the equilibrium
sequences given in these figures are of the supramassive equilibrium sequences characterized
by the constant baryon rest mass given by $M_0=2.10M_\odot$. In Figs. 
\ref{fig:Mb-Phi-const-sup} and \ref{fig:Mb-Phi-const-sup-k2}, each curve is labeled by 
its value of $\Phi_{22}$ which is held constant along the equilibrium sequence.  
For the sake of comparison,
the physical quantities for the non-magnetized equilibrium sequences ($\Phi_{22}=0$)
are also given in Figs. \ref{fig:Mb-Phi-const-sup} and \ref{fig:Mb-Phi-const-sup-k2}. 
Numerical values of the global physical quantities, $\rho_c$, $M$, $R_{\text{cir}}$, 
$\Omega$, $J$, $T/|W|$, $H/|W|$, $\bar{e}$, and $B_{\text{max}}$, for some selected 
models are summarized in Tables \ref{tab:Mb-const-sup-rot} and \ref{tab:Mb-const-sup-rot-k2} 
for the cases of $k=1$ and $k=2$, respectively.

In Figs. \ref{fig:Mb-Phi-const-sup} and \ref{fig:Mb-Phi-const-sup-k2}, we show 
the two different types of the supramassive sequences, as argued before. One is 
the equilibrium sequence that does not connect to the non-rotating solutions, 
whose examples correspond to the sequences labeled by $\Phi_{22}=1.42$ 
for the $k=1$ case and by $\Phi_{22}=0.80$ for the $k=2$ case. In this type 
of the equilibrium sequences, the sequences begin at the mass-shedding limits with 
lower central density $\rho_c$ and move to other mass-shedding limits with 
higher central density $\rho_c$ as the angular velocity $\Omega$ increases. 
This feature is the same as that of the supramassive 
equilibrium sequences of non-magnetized stars (see, e.g., Ref.~\cite{Cook:1992}). 
The other is the equilibrium sequence composed of two disconnected sequences that 
begin at two different non-rotating stars and move to the mass-shedding limits as 
the angular velocity increases.  This type of equilibrium sequences does not 
appear in the non-magnetized stars, in which 
all the equilibrium sequences have a single limit of no rotation as long as it exits.  
This type of the equilibrium sequences corresponds to the sequences labeled by 
$\Phi_{22}=2.13$ for the $k=1$ case and by $\Phi_{22}=1.28$ for the $k=2$ case.
The basic behavior of the global physical quantities for the supramassive 
sequences is nearly independent of 
$\Phi_{22}$ and $k$ if we do not consider the equilibrium sequences composed of 
two disconnected sequences.

\begin{figure*}
  \begin{center}
  \vspace*{40pt}
    \begin{tabular}{c}
      \resizebox{120mm}{!}{\includegraphics{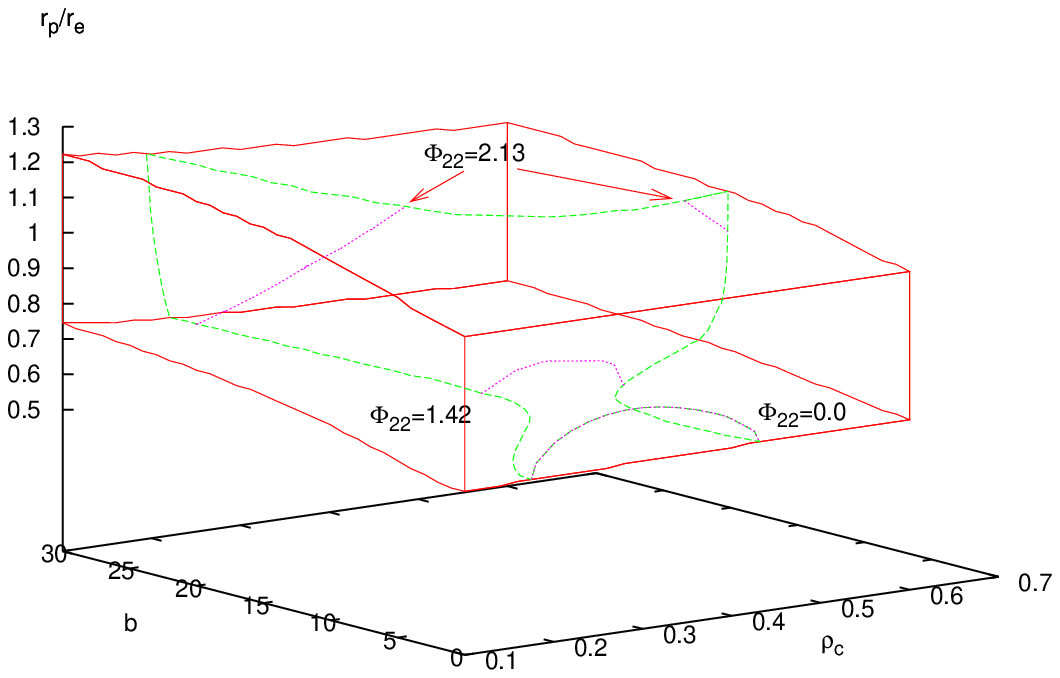}} \\
    \end{tabular}
    \caption{\label{fig:3Dcube-sup}
Same as Fig.~\ref{fig:3Dcube}, but the surface embedded in the cube 
that is drawn with the dashed line boundary shows a set of supramassive equilibrium 
solutions having the same total baryon rest mass. The three dotted curves 
represent the constant magnetic flux sequences referred to these supramassive 
equilibrium solutions.  
     }
  \end{center}
\end{figure*}

In Fig.~\ref{fig:3Dcube-sup}, we show the solution space for the equilibrium 
models of the rotating magnetized stars (similar to Fig. \ref{fig:3Dcube}, in which 
some normal equilibrium sequences are given)  
in order to clearly understand the properties of the supramassive equilibrium sequences.
In Fig.~\ref{fig:3Dcube-sup}, a point inside the cubic region drawn
by the solid lines again corresponds to a physically acceptable solution computed
in this study, 
and the surface drawn with the dashed curve boundary shows a set of 
the supramassive solutions characterized by the baryon rest mass $M_0=2.10M_\odot$. 
In this figure, the three dotted curves indicate the three equilibrium sequences 
of the constant baryon rest mass and 
magnetic flux characterized by $\Phi_{22}=2.13$, $1.42$, and $0$. 
In Fig.~\ref{fig:3Dcube-sup}, 
we see that the two sequences with $\Phi_{22}=0\ {\rm and}\ 1.14$ do not have 
a non-rotating limit and terminate at the mass-shedding limits and that 
the sequence with  $\Phi_{22}=1.71$ is composed of the two disconnected equilibrium 
sequences beginning at the non-rotating limits and terminating at the 
mass-shedding limits. 
(Compare with the plots for the normal sequences in Fig. \ref{fig:3Dcube}).

\begin{table*}
\centering
\begin{minipage}{140mm}
\caption{\label{tab:Mb-const-sup-rot}
Global physical quantities for the supramassive equilibrium sequences 
of the rotating stars with $M_0=2.10M_\odot$ and $k=1$. 
}
\begin{tabular}{ccccccccc}
\hline\hline
$\rho_c[10^{15}{\rm g}/{\rm cm}^3]$ &
$M[M_{\rm \odot}]$     &
$R_{\text{cir}}[{\rm km}]$          &
$\Omega[10^3{\rm rad/s}]$           &
$cJ/G M_\odot^2$                    &
$T/|W|$                             &
$H/|W|$                             &
$\bar{e}$                           &
$B_{\text{max}}[10^{17}{\rm G}]$    \\
\hline\hline
\multicolumn{9}{c}{$\Phi_{22}=0$}\\
\hline
9.817E-01&1.925E+00&1.945E+01&5.965E+00&2.312E+00&9.223E-02&0.000E+00& 2.541E-01&0.000E+00\\
1.348E+00&1.913E+00&1.526E+01&6.383E+00&1.916E+00&6.952E-02&0.000E+00& 1.991E-01&0.000E+00\\
1.797E+00&1.911E+00&1.387E+01&7.221E+00&1.831E+00&6.497E-02&0.000E+00& 1.886E-01&0.000E+00\\
2.359E+00&1.918E+00&1.400E+01&8.775E+00&2.024E+00&7.646E-02&0.000E+00& 2.242E-01&0.000E+00\\
2.421E+00&1.920E+00&1.460E+01&8.955E+00&2.055E+00&7.828E-02&0.000E+00& 2.305E-01&0.000E+00\\
\hline
\multicolumn{9}{c}{$\Phi_{22}=1.42$}\\
\hline
1.692E+00&1.925E+00&2.072E+01&5.289E+00&1.434E+00&4.181E-02&1.076E-01&-8.604E-02&8.365E+00\\
1.797E+00&1.922E+00&1.849E+01&5.374E+00&1.395E+00&3.990E-02&1.053E-01&-8.439E-02&8.668E+00\\
2.247E+00&1.916E+00&1.686E+01&6.014E+00&1.385E+00&3.926E-02&9.679E-02&-5.343E-02&9.797E+00\\
2.327E+00&1.916E+00&1.693E+01&6.185E+00&1.406E+00&4.020E-02&9.542E-02&-4.119E-02&9.985E+00\\
2.461E+00&1.915E+00&1.807E+01&6.462E+00&1.444E+00&4.195E-02&9.325E-02&-2.345E-02&1.027E+01\\
\hline
\multicolumn{9}{c}{$\Phi_{22}=2.13$}\\
\hline
1.342E+00&1.955E+00&2.808E+01&3.336E+00&1.077E+00&2.441E-02&1.940E-01&-5.509E-01&8.626E+00\\
1.470E+00&1.948E+00&2.144E+01&2.977E+00&8.587E-01&1.635E-02&1.896E-01&-6.091E-01&9.183E+00\\
1.557E+00&1.944E+00&1.996E+01&2.649E+00&7.196E-01&1.175E-02&1.866E-01&-6.452E-01&9.539E+00\\
1.700E+00&1.938E+00&1.842E+01&2.035E+00&5.112E-01&6.072E-03&1.819E-01&-6.753E-01&1.010E+01\\
1.937E+00&1.930E+00&1.685E+01&0.000E+00&0.000E+00&0.000E+00&1.744E-01&-6.986E-01&1.095E+01\\
\multicolumn{9}{c}{$\Omega=0$ Limit}\\
2.495E+00&1.917E+00&1.535E+01&0.000E+00&0.000E+00&0.000E+00&1.597E-01&-6.066E-01&1.266E+01\\
2.841E+00&1.914E+00&1.522E+01&2.739E+00&5.283E-01&6.281E-03&1.526E-01&-4.973E-01&1.357E+01\\
3.080E+00&1.913E+00&1.542E+01&3.884E+00&7.404E-01&1.197E-02&1.483E-01&-4.133E-01&1.413E+01\\
3.366E+00&1.913E+00&1.618E+01&5.014E+00&9.548E-01&1.927E-02&1.437E-01&-3.114E-01&1.474E+01\\
3.370E+00&1.913E+00&1.621E+01&5.030E+00&9.579E-01&1.938E-02&1.436E-01&-3.098E-01&1.475E+01\\
\hline
\end{tabular}
\end{minipage}
\end{table*}

\begin{table*}
\centering
\begin{minipage}{140mm}
\caption{\label{tab:Mb-const-sup-rot-k2}
Global physical quantities for the supramassive equilibrium sequences
of the rotating stars with $M_0=2.10M_\odot$ and $k=2$.
}
\begin{tabular}{ccccccccc}
\hline\hline
$\rho_c[10^{15}{\rm g}/{\rm cm}^3]$ &
$M[M_{\rm \odot}]$     &
$R_{\text{cir}}[{\rm km}]$          &
$\Omega[10^3{\rm rad/s}]$           &
$cJ/G M_\odot^2$                    &
$T/|W|$                             &
$H/|W|$                             &
$\bar{e}$                           &
$B_{\text{max}}[10^{17}{\rm G}]$    \\
\hline\hline
\multicolumn{9}{c}{$\Phi_{22}=0.80$}\\
\hline
1.240E+00&1.923E+00&2.053E+01&5.362E+00&1.789E+00&6.045E-02&6.629E-02& 3.648E-02&5.754E+00\\
1.590E+00&1.909E+00&1.607E+01&5.466E+00&1.477E+00&4.441E-02&5.989E-02&-9.328E-03&6.411E+00\\
2.022E+00&1.899E+00&1.461E+01&5.756E+00&1.343E+00&3.771E-02&5.365E-02&-3.089E-02&6.973E+00\\
2.520E+00&1.895E+00&1.417E+01&6.645E+00&1.417E+00&4.118E-02&4.831E-02&-8.473E-03&7.399E+00\\
3.038E+00&1.898E+00&1.593E+01&7.783E+00&1.619E+00&5.082E-02&4.427E-02& 6.335E-02&7.682E+00\\
\hline
\multicolumn{9}{c}{$\Phi_{22}=1.28$}\\
\hline
1.227E+00&1.930E+00&1.931E+01&3.946E+00&1.266E+00&3.308E-02&1.202E-01&-3.038E-01&6.928E+00\\
1.367E+00&1.921E+00&1.750E+01&3.535E+00&1.012E+00&2.238E-02&1.158E-01&-3.693E-01&7.296E+00\\
1.558E+00&1.912E+00&1.601E+01&2.758E+00&7.021E-01&1.138E-02&1.101E-01&-4.344E-01&7.725E+00\\
1.601E+00&1.910E+00&1.577E+01&2.596E+00&6.453E-01&9.524E-03&1.089E-01&-4.455E-01&7.807E+00\\
1.833E+00&1.900E+00&1.464E+01&0.000E+00&0.000E+00&0.000E+00&1.026E-01&-5.142E-01&8.226E+00\\
\multicolumn{9}{c}{$\Omega=0$ Limit}\\
2.884E+00&1.879E+00&1.308E+01&0.000E+00&0.000E+00&0.000E+00&8.380E-02&-4.594E-01&9.282E+00\\
2.905E+00&1.879E+00&1.308E+01&1.231E-01&2.363E-02&3.539E-04&8.353E-02&-4.559E-01&9.293E+00\\
3.274E+00&1.878E+00&1.314E+01&3.102E+00&5.531E-01&7.964E-03&7.940E-02&-3.768E-01&9.478E+00\\
3.335E+00&1.878E+00&1.318E+01&3.665E+00&6.491E-01&9.419E-03&7.882E-02&-3.609E-01&9.504E+00\\
\hline
\end{tabular}
\end{minipage}
\end{table*}

%***********************************
\section{Discussion and Summary}\label{sec:summary}
%***********************************

\subsection{Discussion}

Important findings of the magnetic effects on the equilibrium properties in this study
are summarized as follows ; (1) The mean deformation rates $\bar{e}$ for the strongly
magnetized stars are basically negative, which means that the mean matter
distributions are prolate, even when the stars
rotate at the angular velocity of nearly the mass-shedding limits.
(See, the panel (f) of Figs. \ref{fig:Mb-Phi-const-nom} through
\ref{fig:Mb-Phi-const-sup-k2}.) This implies that rapidly rotating stars containing
strong toroidal magnetic fields could wobble due to their prolate matter
distributions and be potential sources of nearly periodic gravitational
waves for large-scale gravitational wave detectors~\cite{Cutler:2002nw}.
(2) The stronger toroidal magnetic fields lead the mass-shedding of the stars
at the lower $\Omega$ or $T/|W|$. (See, Tables \ref{tab:Mb-const-nom-rot} through
\ref{tab:Mb-const-sup-rot-k2}.) This is because the central concentration of
the matter rises as the toroidal magnetic fields increase, as shown in
Fig. \ref{fig:non-rot}. The larger central concentration of the matter
then induces the mass-shedding at the lower $T/|W|$, as shown in the studies on
the non-magnetized rotating stars (see, e.g., Refs.\cite{Komatsu:1989,Cook:1992}).

\begin{figure*}
  \begin{center}
  \vspace*{40pt}
    \begin{tabular}{cc}
      \resizebox{75mm}{!}{\includegraphics{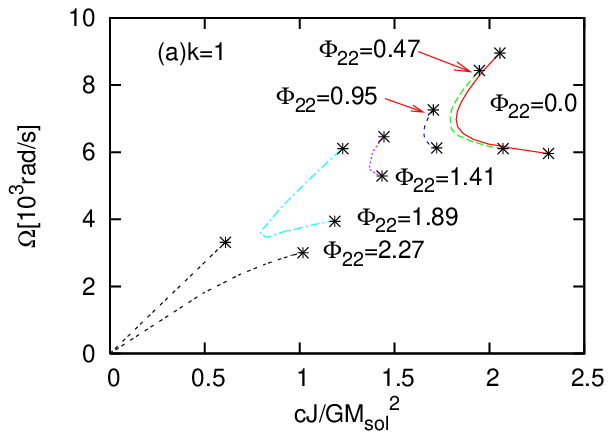}} &
      \resizebox{75mm}{!}{\includegraphics{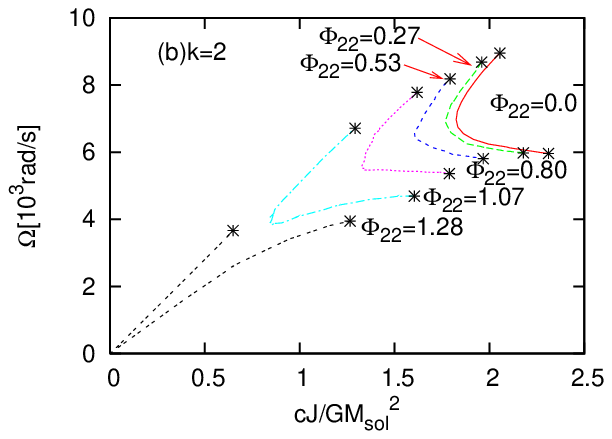}} \\
    \end{tabular}
    \caption{\label{fig:spin-up}
Angular velocity $\Omega$ for the supramassive equilibrium sequence of the rotating
magnetized stars with $M_0=2.10M_\odot$, given as functions of the angular momentum $J$.
(a) $k=1$ and (b) $k=2$. Each curve is labeled by its value of $\Phi_{22}$.
The asterisks indicate the mass-shedding models. 
}
  \end{center}
\end{figure*}
One of the astrophysically important features of the constant baryon mass
equilibrium sequences of the non-magnetized rotating stars is the spin up
of the stars as the stellar angular momentum decreases, found by
Shapiro et al.~\cite{Shapiro:1990} (see, also, Ref.~\cite{Cook:1992}).
This spin up effect of the relativistic stars containing purely poloidal
magnetic fields has also been found in~\cite{Bocquet:1995}. Now, let us investigate
whether this spin up effect can occur for the rotating stars having purely
toroidal magnetic fields, by examining the relationship
between the angular velocity and the angular momentum of the stars.
In Fig.~\ref{fig:spin-up}, we show the angular velocity $\Omega$
as functions of the angular momentum $J$ for the supramassive equilibrium
sequences characterized by the baryon rest mass $M_0=2.10M_\odot$.
In this figure, the panels (a) and (b) give the plots for $k=1$ and $k=2$,
respectively, and each curve is labeled by its value of $\Phi_{22}$, 
which is kept constant along the equilibrium sequence. 
In Fig.~\ref{fig:spin-up}, the plots for the cases of $\Phi_{22}=0$
(the non-magnetized cases) are also shown. From these plots, we confirm that 
for the non-magnetized cases, 
the angular velocities increase as the angular momentum decreases, as shown
in \cite{Cook:1992}.
It can be also seen from this figure that similar spin up effects indeed occur
for the equilibrium sequences characterized by $\Phi_{22}=0.47$, $0.95,$
and $1.41$ for the $k=1$ case and by $\Phi_{22}=0.27$, $0.53,$ for
the $k=2$ case. The two equilibrium sequences with $\Phi_{22}=2.27$
for the $k=1$ case and with $\Phi_{22}=1.28$ for the $k=2$ case are special
in the sense that these sequences have a cusp at the point of
$(J,\ \Omega)=(0,\ 0)$. The one branch of these equilibrium sequences
is seemingly continuous to the other branch.
But, the two equilibrium stars having $(J,\ \Omega)=(0,\ 0)$ are totally
different solutions as discussed before (see, also,
Figs. \ref{fig:Mb-Phi-const-sup} and \ref{fig:Mb-Phi-const-sup-k2}).
Thus, we have to treat these two branches as different equilibrium sequences
because they are discontinuous at $(J,\ \Omega)=(0,\ 0)$. As a result,
there is no spin up effect for such equilibrium sequences.
For the other equilibrium sequences given in Fig.~\ref{fig:spin-up},
on the other hand,
we see that the star initially spins down as it loses angular momentum,
and then it reaches the region of the spin up effects eventually.

Although, in this study, we employ the polytrope equation of state and
simple functional forms of $K$, which determines the magnetic field
distributions inside the star, the numerical scheme in this paper
can be easily extended to treat more complicated functions for $K$ and
realistic equations of state. To perform more quantitative investigations,
we plan to explore the effects of realistic equations of state on
the magnetized relativistic stars.

In this study, we assume the magnetic fields inside the star to be purely
toroidal because we are concerned with the neutron stars whose the toroidal
magnetic fields are much higher than the poloidal fields, which are likely to be
produced in the core-collapse supernova, as suggested in some MHD simulations.
However, this assumption of the purely toroidal magnetic fields oversimplifies
the structures of the magnetized neutron stars because there will be poloidal
magnetic fields in the neutron stars for the following two reasons ;
(1) Seed poloidal magnetic fields and differential rotation are necessary to
amplify the toroidal magnetic fields via the winding up mechanism. (2) It is likely
that the neutron stars observed so far have poloidal magnetic fields, which will
yield the spin down of the neutron stars via electromagnetic radiation.
Therefore, we need to extend the present study
to incorporate the effects of the poloidal magnetic fields in order to
construct more realistic models of the magnetized neutron stars, which 
will be a very hard task. Note that similar extension within
the framework of Newtonian dynamics has already been achieved
\cite{Tomimura:2005,Yoshida:2006a,Yoshida:2006b}.
The prime difficulty encountered is related to the treatment of the spacetime
around the stars containing the mixed poloidal-toroidal magnetic fields.
As discussed in \cite{Gourgoulhon:1993}, we cannot employ the simple
metric (\ref{eq:metric}) if the mixed poloidal-toroidal magnetic fields are
considered. Instead, we have to use a metric whose non-zero components are
increased in number and solve more complicated Einstein equations than those
of the present study, which makes the problem intractable.  
As for the stability of the stars obtained here, which is beyond
the scope of this article, some models obtained in this study might be unstable because
the toroidal magnetic field with $k=1$ induces the kink instability near
the magnetic axes as shown in the Newtonian analysis~\cite{Tayler: 1973}.
However, such instabilities are insignificant for the neutron star models
if the growing timescales of the instabilities are much longer than
the life-time of the stars. Therefore, we have to investigate the detailed
properties of the stability of the stars
by using linear perturbation analysis or numerical simulations to
see whether or not such instabilities are indeed effective for the neutron
star models. The two problems concerning the magnetized star models mentioned
above, however, remain as future work of the greatest difficulty.

\subsection{Summary}

In this study, we have investigated the effects of the purely toroidal magnetic 
field on the equilibrium structures of the relativistic stars. 
The master equations for obtaining equilibrium solutions of relativistic 
rotating stars containing purely toroidal magnetic fields have been derived 
for the first time. In our formalism, the distribution of the magnetic fields 
are determined by an arbitrary function of $\rho_0 h g_B$, $K(\rho_0 h g_B)$, 
which can be chosen freely as long as the boundary conditions for 
the magnetic fields on the magnetic axis and the surface of the star are 
satisfied. To solve these master equations numerically, we have 
extended the Cook-Shapiro-Teukolsky scheme for calculating relativistic rotating 
stars containing no magnetic field to incorporate the effects of the purely 
toroidal magnetic fields. By using the numerical scheme, we have then calculated  
a large number of the equilibrium configurations for particular function form of 
$K$ in order to explore the equilibrium properties. 
We have also constructed the equilibrium sequences of the constant baryon mass 
and/or the constant magnetic flux, which model the evolution of an isolated neutron 
star as it loses angular momentum via gravitational waves. 
Important properties of the equilibrium configurations of the magnetized stars 
obtained in this study are summarized as follows ; (1) For the non-rotating stars, 
the matter distribution of the stars is prolately distorted due to 
the toroidal magnetic fields. (2) For the rapidly rotating 
stars, on the other hand, the shape of the stellar surface becomes oblate because 
of the centrifugal force. But, the matter distribution deep inside the star is 
sufficiently prolate for the mean deformation rate $\bar{e}$ to be negative. 
Here, a negative value of $\bar{e}$ means that mean matter distribution is prolate.   
(3) The stronger toroidal magnetic fields lead to the mass-shedding of the stars
at the lower $\Omega$ or $T/|W|$.
(4) For some supramassive equilibrium sequences of the constant baryon mass and 
magnetic flux, the stars can spin up as they are losing angular momentum.

%*************************
\section*{Acknowledgments}
%*************************
We thank K.~i. Maeda, Y. Eriguchi, S. Yamada, K. Kotake, and Y. Sekiguchi 
for informative discussions. K.K. is supported by Japan Society for
Promotion of Science (JSPS) Research Fellowships. This work was partly 
supported by a Grant-in-Aid for Scientific Research (C) from the Japan 
Society for the Promotion of Science (1954039).

%**************************

%********************
%********
\appendix
%********
%*******************************************************
\section{Magnetic energy and flux for the star 
with purely toroidal magnetic fields}\label{sec:apdixA}
%*******************************************************
For the perfectly conductive fluid,  
the total energy-momentum four-vector $p^\mu$ measured by the observers with  
the matter four-velocity $u^\mu$ is given by 
%------------------------------------------------------------------------
\begin{eqnarray}
p^\mu&=&{T^\mu}_{\nu} u^\nu \nonumber \\
&=&-\left(\rho_0+\rho_0 e+\frac{1}{8\pi}\,B^\alpha B_\alpha\right) u^\mu \,. 
\end{eqnarray}
%-------------------------------------------------------------------------
Then, the total proper energy of the matter $E_{\rm proper}$ may be defined as  
%------------------------------------------------------------------------
\begin{eqnarray}
E_{\rm proper}&=&-\int p^\mu d^3\Sigma_\mu \nonumber \\
&=&\int \left(\rho_0+\rho_0 e+\frac{1}{8\pi}\,B^\alpha B_\alpha\right) 
u^\mu d^3\Sigma_\mu \nonumber \\
&=&\int \left(\rho_0+\rho_0 e+\frac{1}{8\pi}\,B^\alpha B_\alpha\right) 
u^t \sqrt{-g}\, d^3x \,.
\label{defEprop}
\end{eqnarray}
%-------------------------------------------------------------------------
We see that the first, the second, and the third term in the last line  
of equation (\ref{defEprop}) respectively represent the total baryon rest 
mass, the total internal energy, and the magnetic energy of the system. 
We may therefore define the stellar magnetic energy $H$ as 
%------------------------------------------------------------------------
\begin{eqnarray}
H=\frac{1}{8\pi}\,\int B^\alpha B_\alpha u^t \sqrt{-g}\, d^3x \,.
\label{defH}
\end{eqnarray}
%-------------------------------------------------------------------------

Next, let us consider the magnetic flux. Since the two conditions, given by 
%------------------------------------------------------------------------
\begin{eqnarray}
&&u^\mu F_{[\mu\nu,\alpha]}=0\,, \\
&& F_{\mu\nu}\,u^\nu=0\,, 
\end{eqnarray}
%-------------------------------------------------------------------------  
are satisfied because of the Maxwell equation and infinite conductivity of 
the fluid, we have 
%------------------------------------------------------------------------
\begin{eqnarray}
{\cal L}_{\alpha u} F_{\mu\nu} = 0\,, 
\end{eqnarray}
%-------------------------------------------------------------------------
where ${\cal L}_{u}$ stands for the Lie differentiation with respect to 
$u^\mu$, and $\alpha$ denotes an arbitrary function. This implies that 
the flux of $F_{\mu\nu}$ is conserved along any tube generated by curves  
parallel to $u^\mu$. In other words, the integral of $F_{\mu\nu}$, 
defined as 
%------------------------------------------------------------------------
\begin{eqnarray}
\Phi=\int_{\Sigma(\tau)}  F_{\mu\nu}\, dx^\mu \wedge dx^\nu \,, 
\end{eqnarray}
%------------------------------------------------------------------------- 
where $\Sigma(\tau)$ is a two-dimensional surface co-moving with the fluid 
flow parametrized by its proper time $\tau$, is conserved in the sense 
that 
%------------------------------------------------------------------------
\begin{eqnarray}
{d \Phi\over d\tau}=0 \,. 
\end{eqnarray}
%-------------------------------------------------------------------------
In the present situation, we can take as $\Sigma(\tau)$ the two-surface 
orthogonal to the plane defined by the two Killing vectors parametrized 
by $r$ and $\theta$. For the present situation, we can then reduce $\Phi$ into 
%------------------------------------------------------------------------
\begin{eqnarray}
\Phi=\int^\infty_0 dr \int^\pi_0 d\theta\, F_{12}\,, 
\end{eqnarray}
%-------------------------------------------------------------------------
which is frequently called the magnetic flux.  


\begin{thebibliography}{99}
%**************************

\bibitem{Bocquet:1995} 
  Bocquet,~M., Bonazzola,~S., Gourgoulhon,~E., \& Novak,~J. 1995,
  A\&A,  301, 757

\bibitem{Bonazzola:1994}
  Bonazzola,~S. \& Gourgoulhon,~E., 1994, 
  CQG, 11, 1775

\bibitem{Bonazzola:1995rb}
  Bonazzola,~S., \& Gourgoulhon,~E., 1996,
  %``Gravitational waves from pulsars: emission by the magnetic field induced
  %distortion,''
  A\&A  312, 675
  %[arXiv:astro-ph/9602107].
  %%CITATION = AAEJA,312,675;%%

%\bibitem{Braithwaite:2004}
%  Braithwaite,~J. \& Spruit,~H.~C. 2004,
%  Nature, 431, 819

%\bibitem{Braithwaite:2006}
%  Braithwaite,~J. \& Spruit,~H.~C. 2006,
%  A\&A,  450, 1097

\bibitem{Cardall:2001}
 Cardall,~C.~Y., Prakash,~M., \& Lattimer,~J.~M. 2001 
 ApJ, 554, 322 

\bibitem{Carter:1969}
  Carter,~B., 1969, 
  J.\ Math.\ Phys.\ 10, 70

\bibitem{Chandra:1953}
  Chandrasekhar,~S., \& Fermi,~E. 1953,
  ApJ  118, 116

\bibitem{Cook:1992}
  Cook,~G.~B., Shapiro,S.~L. \& Teukolsky,~S.~A., 
  ApJ 398, 203 (1992), 422, 227, (1994)

\bibitem{Cutler:2002nw}
  Cutler,~C., 2002, 
  %``Gravitational waves from neutron stars with large toroidal B-fields,''
  PRD 66, 084025
  %[arXiv:gr-qc/0206051].
  %%CITATION = PHRVA,D66,084025;%%

\bibitem{Ferrario:2007bt}
  Ferrario,~L. \& Wickramasinghe,~D., 2007
  %``The birth properties of Galactic millisecond radio pulsars,''
  MNRAS, 375, 1009
  %[arXiv:astro-ph/0701444].
  %%CITATION = MNRAA,375,1009;%%

\bibitem{Geppert:2006cp}
  Geppert,~U. and Rheinhardt,~M.
  %``Magnetars versus Radio Pulsars: MHD Stability in Newborn Highly Magnetized
  %Neutron Stars,''
  A\& A, 456, 639
  %arXiv:astro-ph/0606120.
  %%CITATION = ASTRO-PH/0606120;%%

\bibitem{Gourgoulhon:1993}
  Gourgoulhon,~E., \& Bonazzola,~S., 1993, 
  PRD, 48, 2635 

\bibitem{Gourgoulhon:1994}
  Gourgoulhon,~E., \& Bonazzola,~S., 1994
  CQG,  11, 443

\bibitem{Harding:2006} 
  Harding, A.~K., \& Lai, D.\ 2006, 
  Reports of Progress in Physics, 69, 2631 

\bibitem{Hachisu:1986}
  Hachisu, I. 1986, 
  ApJS, 61, 479 

\bibitem{Ioka:2003}
  Ioka,~K. \& Sasaki,~M. 2003,
  PRD  67, 124026

\bibitem{Ioka:2004}
  Ioka,~K. \& Sasaki,~M. 2004,
  ApJ   600, 296

\bibitem{Kiuchi:2007pa}
  Kiuchi,~K., \& Kotake,~K.,
  %``Equilibrium Configurations of Strongly Magnetized Neutron Stars with
  %Realistic Equations of State,''
  appeared in MNRAS, arXiv:0708.3597 [astro-ph].
  %%CITATION = ARXIV:0708.3597;%%

\bibitem{Komatsu:1989}
  Komatsu,~H., Eriguchi,~Y., \& Hachisu,~I., 
  MNRAS, 237, 355 (1989), 239, 153 (1989)

\bibitem{Konno:1999zv}
  Konno,~K., Obata,~T.\& Kojima,~Y. 1999,
  %``Deformation of relativistic magnetized stars,''
  A\&A, 352, 211
  %[arXiv:gr-qc/9910038].
  %%CITATION = AAEJA,352,211;%%

\bibitem{Kotake:2004} 
  Kotake, K., Sawai, H., Yamada, S., \& Sato, K.\ 2004, ApJ, 608, 391 

\bibitem{Kotake:2006} 
  Kotake, K., Sato, K., \& Takahashi, K.\ 2006, 
  Reports of Progress in Physics, 69, 971 

\bibitem{kouve} 
  Kouveliotou, C., et al.\ 1998, Nature, 393, 235 

\bibitem{Lattimer:2006}
  Lattimer,~J.~M. \& Prakash,~M. 2006
  astro-ph/0612440.

\bibitem{Livne:2007} 
  Livne, E., Dessart, L., Burrows, A., \& Meakin, C.~A.\ 2007, 
  ApJS, 170, 187 

%\bibitem{Markey:1973}
%  Markey,~P., \& Taylar.~R.~J. 1973
%  MNRAS, 163, 77

%\bibitem{Markey:1974}
%  Markey,~P., \& Taylar.~R.~J. 1974
%  MNRAS, 168, 505

\bibitem{Miketinac:1973}
  Miketinac,~M.~J. 1973
  Ap\&SS, 22, 413

\bibitem{Moissenko:2006} 
  Moiseenko, S.~G., Bisnovatyi-Kogan, G.~S., \& Ardeljan, N.~V.\ 2006, 
  MNRAS, 370, 501 

\bibitem{Obergaulinger:2006}
  Obergaulinger, M., Aloy, M.~A., M\"{u}ller, E.\ 2006, 
  A\&A, 450, 1107 

\bibitem{Oron: 2002}
  Oron,~A., 2002, 
  PRD, 66, 023006

%\bibitem{Salgado:1994} 
%  Salgado,~M., Bonazzola,~S., Gourgoulhon,~E., \& Haensel,~P. 1994,
%  A\&A, 291, 155

\bibitem{Shapiro:1990}
  Shapiro,~S.~L., Teukolsky,~S.A., \& Nakamura,~T. 1990, 
  ApJ, 357, L17 

\bibitem{Shibata:2005ss}
  Shibata,~M., Taniguchi,~K., \& Uryu,~K., 2005, 
  %``Merger of binary neutron stars with realistic equations of state in  full
  %general relativity,''
  PRD 71, 084021
  %[arXiv:gr-qc/0503119].
  %%CITATION = PHRVA,D71,084021;%%

\bibitem{Shibata:2006hr}
  Shibata,~M., Liu,~Y.~T., Shapiro,~S.~L., \& Stephens,~B.~C. 2006
  %``Magnetorotational collapse of massive stellar cores to neutron stars:
  %Simulations in full general relativity,''
  PRD 74, 104026
  %[arXiv:astro-ph/0610840].
  %%CITATION = PHRVA,D74,104026;%%
  
\bibitem{Tayler: 1973}
  Tayler,~R.~J., 1973
  MNRAS, 161, 365

\bibitem{Thompson:1993hn}
  Thompson,~C. \& Duncan,~R.~C. 1993,
  %``Neutron star dynamos and the origins of pulsar magnetism,''
  ApJ, 408, 194
  %%CITATION = ASJOA,408,194;%%

\bibitem{Thompson:1995gw}
  Thompson,~C. \& Duncan,~R.~C. 1995,
  %``The Soft gamma repeaters as very strongly magnetized neutron stars - 1.
  %Radiative mechanism for outbursts,''
  MNRAS,  275, 255
  %%CITATION = MNRAA,275,255;%%

\bibitem{Thompson:1996pe}
  Thompson,~C. \& Duncan,~R.~C. 1996,
  %``The Soft gamma repeaters as very strongly magnetized neutron stars. 2.
  %Quiescent neutrino, x-ray, and Alfven wave emission,''
  ApJ 473, 322
  %%CITATION = ASJOA,473,322;%%

\bibitem{Tomimura:2005}
  Tomiumra,~Y. \& Eriguchi,~Y., 2005 
  MNRAS,  359, 1117

\bibitem{trehan1972}
  Trehan,~S.~K. \& Uberoi,~M.~S. 1972
  ApJ, 175, 161 

\bibitem{Wald:1984}
  Wald,~R.~M., 1984, 
  {\it General Relativity} (The University of Chicago Press),

\bibitem{Yoshida:2006a}
%\bibitem[\protect\citeauthoryear{Yoshida \& Eriguchi}{2006}]{yos}
  Yoshida,~S. \& Eriguchi,~Y. 2006,
  ApJS,  164, 156

\bibitem{Yoshida:2006b}
%\bibitem[\protect\citeauthoryear{Yoshida et al.}{2006}]{yos2} 
  Yoshida,~S., Yoshida,~S., \& Eriguchi,~Y. 2006,
  ApJ,  651, 462

\bibitem{anna}
  Watts, A.\ 2006, 36th COSPAR Scientific Assembly, 36, 168 

\bibitem{wod} 
  Woods,~P.~M. \& Thompson,~C. 2004,
  %``Soft Gamma Repeaters and Anomalous X-ray Pulsars: Magnetar Candidates,''
  arXiv:astro-ph/0406133.

%\bibitem{Wright:1973}
%  Wright,~G.~A.~E. 1973,
%  MNRAS,  162, 339

\bibitem{Yamada:2004} 
  Yamada, S., \& Sawai, H.\ 2004, 
  ApJ, 608, 907 

%********************
\end{thebibliography}
\end{document}